\documentclass{article}
\usepackage[utf8]{inputenc}
\usepackage{amsmath,commath,amssymb,mathtools}
\usepackage{bm}
\usepackage[colorlinks,citecolor=cyan]{hyperref}
\usepackage[capitalize]{cleveref}
\usepackage{physics}
\usepackage{algorithm,algorithmicx,algpseudocode}
\usepackage{csvsimple}
\usepackage[usenames,dvipsnames]{xcolor}
\usepackage{url}
\usepackage{scalerel,stackengine}
\usepackage{siunitx,booktabs}
\usepackage[left=2.5cm,right=2.5cm,top=3cm,bottom=3cm]{geometry}
\usepackage{natbib}
\usepackage{fancyhdr}
\setcitestyle{numbers}

\stackMath
\newcommand\reallywidehat[1]{%
\savestack{\tmpbox}{\stretchto{%
  \scaleto{%
    \scalerel*[\widthof{\ensuremath{#1}}]{\kern.1pt\mathchar"0362\kern.1pt}%
    {\rule{0ex}{\textheight}}%WIDTH-LIMITED CIRCUMFLEX
  }{\textheight}% 
}{2.4ex}}%
\stackon[-6.9pt]{#1}{\tmpbox}%
}

\newcommand{\appropto}{\mathrel{\vcenter{
			\offinterlineskip\halign{\hfil$##$\cr
				\propto\cr\noalign{\kern2pt}\sim\cr\noalign{\kern-2pt}}}}}

\def \Ra {\textrm{Ra}}
\def \Nu {\textrm{Nu}}

\fancypagestyle{cc}{%
	\fancyhf{}
	
	\fancyfoot[C]{© 2020. This manuscript version is made available under the CC-BY-NC-ND 4.0 license \url{http://creativecommons.org/licenses/by-nc-nd/4.0/}.}
}

\title{Fourier Continuation method for incompressible fluids with boundaries}
\author{M.~Fontana$^1$\thanks{mfontana@df.uba.ar}~, O.P.~Bruno$^2$, P.D.~Mininni$^1$, and P.~Dmitruk$^1$}
\date{$^1$\small{Universidad de Buenos Aires, Facultad de Ciencias Exactas y Naturales, Departamento de Física, \& IFIBA, CONICET, Ciudad Universitaria, Buenos Aires 1428, Argentina.}\\
$^2$\small{Computing and Mathematical Sciences, Caltech, Pasadena, CA 91125, USA.}}

\begin{document}
\maketitle

\thispagestyle{cc}
\begin{abstract}
We present a Fourier Continuation-based parallel pseudospectral method for incompressible fluids in  cuboid non-periodic domains. The method produces dispersionless and dissipationless derivatives with fast spectral convergence inside  the  domain, and with very high order convergence at the boundaries. Incompressibility is imposed by solving a Poisson equation for the pressure. Being Fourier-based, the method allows for fast computation of spectral transforms. It is compatible with  uniform grids (although refined or nested meshes can also be implemented), which in turn allows for  explicit time integration at sufficiently high Reynolds numbers. Using a new parallel code named \textsc{SPECTER} we illustrate the method with two problems: channel flow, and plane Rayleigh-Bénard convection under the Boussinesq approximation. In both cases the method yields results compatible with previous studies using other high-order numerical methods, with mild requirements on the time step for stability.
\end{abstract}

\maketitle

\section{Introduction}

Numerically solving the Navier-Stokes equations is one of the standard tools employed for researching the behavior of turbulent flows. As the study of turbulence often requires quantification of high-order statistical quantities, high-order numerical methods are in many cases preferred \cite{Yeung_2015, Khurshid_2018}. Moreover, it is believed that at large enough distances from physical boundaries some properties of turbulent flows are universal, and thus studies sometimes concentrate on the dynamics of the bulk of the fluid. This motivates the study of ``isotropic and homogeneous turbulence," a problem which can be successfully modeled employing periodic boundary conditions. This comes with great computational advantages, as the Navier-Stokes equations can be very efficiently solved in periodic domains utilizing Fourier representations and pseudospectral calculations \cite{Orszag1969, Orszag1972, Canuto1988, Fornberg1998}. Moreover, if compressibility effects can also be neglected, enforcing the continuity condition in this case reduces to solving a Poisson equation for the fluid pressure with periodic boundary conditions, which in the wavenumber domain can be easily and efficiently accomplished.

Notwithstanding the major importance of understanding bulk dynamics in turbulent flows, the statistics of turbulence near boundaries is clearly as important, with implications for industrial, geophysical, environmental, and astrophysical flows \cite{Pope2000, Hirsch2007}. Classical problems where actual physical boundaries must be considered include the flow through a pipe or a channel maintained either by an imposed pressure difference \cite{Kim1987, Zagarola1998, Eckhardt2007, Avila2011, Lee2015} or a moving wall \cite{Orszag1980, Salewski2015, Klotz2017}, and the natural convection occurring when a box is heated at one end and cooled at the other \cite{Biringen1990, Bodenschatz_2000, Amati2005, Lohse_2010}. Also, in the case of conducting fluids, Hartmann flows represent a classical example of wall-bounded magnetohydrodynamic (MHD) turbulence \cite{Moresco2004, Krasnov2012}.

However, when the presence of walls must be accounted for, the classical pseudospectral method using fast Fourier transforms (FFTs) as described in \cite{Orszag1969} cannot be employed, as the Gibbs phenomenon severely degrades (or even forbids) convergence. Consequently, several other high-order representations for the non-periodic directions have been introduced \cite{Lele1992}, including fully spectral methods \cite{Gottlieb1977, Shan_1991, Mininni_2006, Fontana_2018}, Chebyshev pseudospectral methods \cite{Canuto1988, Fornberg1998, Julien_2009}, B-splines \cite{Botella_2003,Kwok2001}, and spectral element methods \cite{Patera_1984, Rosenberg_2006, Rosenberg2007}. All these techniques have been successfully employed in several scientific solvers for turbulent flows and for other partial differential equations (PDEs). When dealing with incompressible flows, however, directly solving the Poisson equation for the pressure becomes a very computationally demanding task, and can even be ill-behaved for some methods (see, e.g., discussions in \cite{Fornberg1998, Julien_2009}). Common strategies to avoid this problem include recasting the equations to another set of variables that automatically enforce incompressibility such as the normal velocity-normal vorticity formulation \cite{Kim1987}, using  preconditioned methods \cite{Deville2002}, or solving the Poisson equation employing methods based on Green functions and integral equations \cite{Ying2006,Bruno2012,Bruno2018}.

Another approach for simulating incompressible fluids with boundaries while retaining a Fourier representation of the fields is the usage of penalization techniques, as in virtual or immersed boundary methods \cite{Peskin1972, Iaccarino2003, Wirth2005, Kolomenskiy2009}. In these methods, non-physical extra terms are added to the equations in order to impose the boundary conditions. This has the drawback that near the boundaries the order of the approximation is notably lowered, resulting even in slow algebraic global convergence of the solutions in the entire domain (see, e.g., the discussion in p.~205 of Ref.~\cite{Canuto2007}). More recently, a high-order solver entirely based on Fourier representations was presented in \cite{Albin2011, Bruno2016} for compressible flows with non-periodic boundary conditions. In that work, an efficient and high-order Fourier Continuation with Gram polynomials (FC-Gram) technique \cite{Lyon2010a, Lyon2011} was employed to circumvent the Gibbs phenomenon, and boundary conditions were enforced simply by strong imposition of the conditions in physical space.

In this paper we present an FC-Gram based method for incompressible flows in cuboid non-periodic domains, that produces dispersionless derivatives with spectral accuracy inside the domain, and with very high order convergence at the boundaries. The usage of Fourier basis allows for the introduction of an effective Poisson solver for the pressure that is numerically well behaved and that satisfies the divergence-free condition of the velocity field, and its boundary condition, with high accuracy. Moreover, being Fourier-based, the method has the advantage of being compatible with uniform grids, which in turn allows for efficient explicit time integration without requirements of small time steps for stability when considering flows at high Reynolds numbers. After presenting the method, we validate a three-dimensional numerical implementation in two paradigmatic problems: channel flow, and plane Rayleigh-B\'enard convection under the Boussinesq approximation.

The remaining of the text is organized as follows: In Section \ref{sec:equations} we introduce the governing equations and the notational conventions. In Section \ref{sec:FCGram} we give a brief summary of the key ideas and advantages of the FC-Gram technique. In Section \ref{sec:method}, the proposed numerical method for the incompressible Navier-Stokes equations is presented, including the method to solve for the pressure, as well as low- and high-order time stepping schemes. A parallelization method that scales well with the numerical method presented is discussed in Section \ref{sec:numerics}. While the method is described in detail for the incompressible Navier-Stokes equations, generalizations to other PDEs with solenoidal vector fields, such as the incompressible MHD equations, or the incompressible Boussinesq equations, are straightforward, as illustrated with examples in the following sections. In Section \ref{sec:chan}, turbulent plane Poiseuille flow simulations are presented as an example to validate our algorithm, while in Section \ref{sec:Rayleigh} plane Rayleigh-Benard convection simulations are discussed. Finally, section \ref{sec:conclusion} presents our conclusions.
\section{Governing equations}
\label{sec:equations}

\begin{figure}[t]
	\centering
	\includegraphics[width=.75\textwidth, keepaspectratio=true]{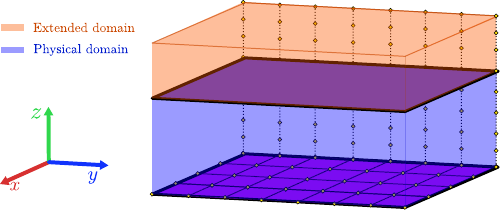}
	\caption{Schematic representation of the domain, with the physical domain in blue, the extended domain used for the FC-Gram method in orange, and the Cartesian axes indicated for convenience. The two horizontal (darker) lids indicate the no-slip boundaries in $z$. Boundaries in the $x$ and $y$ directions are periodic. A few grid points are shown as a reference.}
	\label{met:fig:domain}
\end{figure}

In the simplest configuration, we want to solve the three-dimensional (3D) incompressible forced Navier-Stokes equations
\begin{align}
    \label{equations:eq:navier-stokes}
    \dpd{\bm{v}}{t} + (\bm{v} \cdot \bm{\nabla}) \bm{v} &= - \bm \nabla p + \nu \nabla^2 \bm{v} +\bm{f},\\
    \bm \nabla \cdot \bm v &= 0,
    \label{equations:eq:incompressibility}
\end{align}
in a $(0,0,0)\times(L_x,L_y,L_z)$ domain, over a uniform grid of $N_x \times N_y \times N_z$ points (see Fig.~\ref{met:fig:domain}). In Eq.~\eqref{equations:eq:navier-stokes} $\bm v$ is the velocity field, $p$ is the pressure, $\bm f$ is a solenoidal forcing field, $\nu$ is the kinematic viscosity, and the fluid mass density $\rho$ is assumed to be equal to 1 (in dimensionless units) for simplicity. We are particularly interested in the moderate and low kinematic viscosity cases, in which turbulent behavior takes place. In these cases the Courant–Friedrichs–Lewy (CFL) constraint is dominated by the advection term in Eq.~\eqref{equations:eq:navier-stokes}, and explicit time stepping is, when allowed by the spatial discretization, the time integration method of choice.

For the velocity field we assume periodic boundary conditions in the $x$ and $y$ directions, and no-slip boundary conditions in the $z$ direction,
\begin{align}
    \left. \bm{v} \right|_{x=0} &= \left. \bm{v} \right|_{x=L_x},\\
    \left. \bm{v} \right|_{y=0} &= \left. \bm{v} \right|_{y=L_y},\\
    \left. \bm{v} \right|_{z=0} &= \left. \bm{v} \right|_{z=L_z} = \bm 0.
    \label{equations:eq:boundary_conditions}
\end{align}
The periodic boundary conditions in $x$ and $y$ also imply that all derivatives of the velocity field $\bm v$ must be periodic,
\begin{align}
    \partial_x^{n} \bm v |_{x=0} &=  \partial_x^{n} \bm v |_{x=L_x} \qquad \forall n \in \mathbb N,\\
    \partial_y^{n} \bm v |_{y=0} &=  \partial_y^{n} \bm v |_{y=L_y} \qquad \forall n \in \mathbb N.
    \label{equations:eq:boundary_conditions_der}
\end{align}

Before proceeding, note the choice made here of the PDE to be solved, as well as of the boundary conditions, is done for the sake of clarity in the presentation of the method. Other PDEs involving constrains in the fields being solenoidal (as, e.g., the MHD equations for which the magnetic field $\bm B$ satisfies the condition $\bm \nabla \cdot \bm B = 0$), or other boundary conditions for the fields, can be implemented using the method described in Sections \ref{sec:FCGram} and \ref{sec:method}. In the same spirit, a single non-periodic direction is employed in this work both for clarity and for validation with the extensively researched physical problems considered in Sections \ref{sec:chan} and \ref{sec:Rayleigh}. However, the generalization of our method to two or more non-periodic directions is possible.

By taking the divergence of \cref{equations:eq:navier-stokes} and utilizing \cref{equations:eq:incompressibility}, one gets a Poisson equation for the pressure 
\begin{equation}
\label{equations:eq:poisson}
\nabla^2 p = - \bm \nabla \cdot \left[ \left( \bm v \cdot \bm \nabla \right) \bm v \right].
\end{equation}
This implies that the pressure gradient is not an independent variable for incompressible flows, but instead acts as a Lagrange multiplier restricting the velocity field to the subspace of solenoidal fields. However, replacing \cref{equations:eq:incompressibility} with \cref{equations:eq:poisson} makes it evident that appropriate boundary conditions compatible with the no-slip condition for $\bm v$ must be supplemented to the pressure. The most natural approach is to project \cref{equations:eq:navier-stokes} at the boundaries in the direction normal or tangential to the wall, and to solve either a Neumann or a Dirichlet problem. There is no {\it a priori} reason to assume that both approaches lead to the same solution, a topic which has been of wide discussion and will not be treated here. In the same line of thinking, independent boundary conditions for the pressure could be imposed; we defer the study of this subject for future work. For references on these topics see for example \cite{Orszag1974,Rempfer2006,Kress2000} and references therein, and the discussion in pp.~83-87 of Ref.~\cite{Gallavotti}. For this work, we project \cref{equations:eq:navier-stokes} in the wall normal direction and solve \cref{equations:eq:poisson} for the pressure employing Neumann boundary conditions, as discussed in \cite{Orszag1986}.

Finally, it is worth mentioning that Eqs.~\eqref{equations:eq:navier-stokes}, \eqref{equations:eq:incompressibility}, and \eqref{equations:eq:boundary_conditions} can be written in several equivalent formulations, such as the normal velocity-normal vorticity formulation employed, for example, in the landmark publication by Kim, Moin and Moser \cite{Kim1987}. However, both for simplicity and scalability to other sets of equations (as, for example,  Boussinesq or MHD problems), the standard Navier-Stokes formulation in terms of $\bm v$ and $p$ is chosen in this work.

\section{FC-Gram transformation method}
\label{sec:FCGram}

For the spatial discretization in the periodic $x$ and $y$ directions we follow the ideas presented in \cite{Orszag1969} which are now standard practice in pseudospectral methods for PDEs, in which a trigonometric basis is used for expanding $\bm{v}$, $p$ and $\bm f$ in the $(k_x, k_y)$ domain. Using the same basis for the non-periodic $z$ direction in a straightforward manner would result in the well known Gibbs phenomenon, and produce a major loss of accuracy \cite{Bocher1906}.

\begin{figure}[t]
	\centering
	\includegraphics[width=.65\textwidth, keepaspectratio=true]{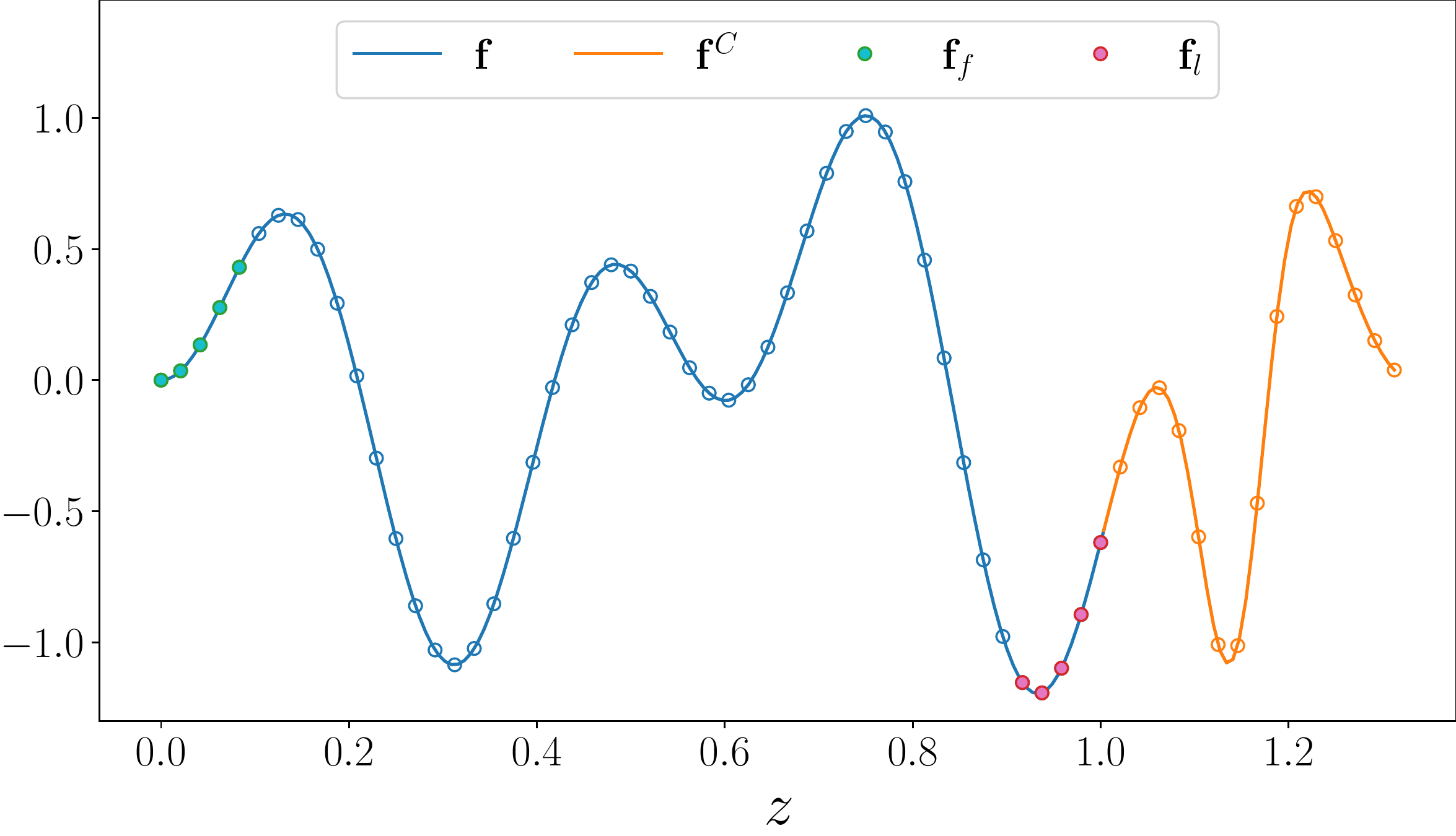}
	\caption{The FC-Gram method applied to the function \mbox{$f(z)=\sin(15z)\sin(5.5z)\exp(0.3z)$}. The original data $\mathbf f$ (blue line) contains 49 equispaced points in the $[0,1]$ interval, and the ``continued" data $\mathbf f^C$ (orange line) consists of 15 values (i.e. the number of continuation points is $C=15$). 5 matching points (i.e. $d=5$) were used on each boundary, highlighted in green and magenta (respectively indicated as $\mathbf f_f$ and $\mathbf f_l$). Note the data in the extended domain is periodic, and thus representable by a Fourier series without Gibbs phenomena.}
	\label{fc:fig:fc_demo}
\end{figure}

The Gibbs ringing problem is circumvented in this work by utilizing a Fourier Continuation technique, in which an extended domain of length $L_z'$ is considered in the $z$ direction (see Fig.~\ref{met:fig:domain}), and an appropriate periodic continuation is computed for all functions in the $[L_z,L_z')$ interval using Gram polynomials \cite{Lyon2010a, Lyon2011, Albin2011, Albin2012} (see an example in Fig.~\ref{fc:fig:fc_demo} for a non-periodic function dependent only on $z$ in the [0,1] interval). The union of the original and continued points can then be Fourier transformed to the $k_z$ space (i.e., expanded into a trigonometric basis). In this way, the error arising from the Gibbs phenomenon is severely reduced. This technique has also the advantage of producing high-order and  dispersionless spatial derivatives \cite{Bruno2014}, a property of major importance when dealing with the large range of scales present in highly turbulent flows.

To describe the method in more detail, let's consider a 3D ``continued" scalar field $\phi^c$ (which can be the pressure field $p$, each Cartesian component of the velocity field $v_i$ with $i=x$, $y$, or $z$, a temperature field, or the Cartesian components of other vector fields when dealing with other PDEs). The continued scalar field $\phi^c$ is defined over the $(0,0,0)\times(L_x,Ly,L_z')$ extended domain, and matches the physical field $\phi$ on the $(0,0,0)\times(L_x,L_y,L_z)$ region. As fields in the extended domain are 3D-periodic, $\phi^c$ can be represented as
\begin{equation}
\phi^c_{qrs} = \sum_{n=m=l=0}^{N_x-1,N_y-1,N_z+C-1} \hat \phi^c_{nml} e^{i \bm{k}_{nml}\cdot \bm x_{qrs}},
\end{equation}
where $\hat \phi^c$ are the Fourier coefficients of $\phi^c$, $\bm x_{qrs}$ is the position vector in the extended domain $(q\Delta x, r\Delta y, s\Delta z)$ (with $\Delta x$, $\Delta y$, and $\Delta z$ the spatial resolution, and $q$, $r$, and $s$ integers that label each point of the spatial grid), $\bm k_{nml}$ is the wavenumber vector
\begin{equation}
    \bm k_{nml} = \frac{\pi(2n-N_x)}{L_x}\bm{\hat x} + \frac{\pi(2m-N_y)}{L_y}\bm{\hat y} + \frac{\pi(2l-N_z-C)}{L_z'}{\bm{\hat z}} ,
    \label{eq:k}
\end{equation}
(assuming an even number of gridpoints), and $C$ is the number of continuation points in the $z$ direction.

As mentioned above, to compute the periodic extension of $\phi(x,y,z)$ in the $z$ direction we employ the FC-Gram method first presented in \cite{Lyon2010a, Albin2011, Albin2012} and employed with great success in \cite{Bruno2014, Bruno2016, Amlani2016, Bruno2019}. The main idea behind the FC-Gram method is to project the values near the boundaries onto an orthogonal polynomial basis, and to calculate appropriate continuations for each element of the basis. As this is basically a one-dimensional (1D) problem in each direction in which quantities must be extended (i.e., in each direction without periodic boundary conditions), we consider a non-periodic scalar function $f(z)$ (not to be confused with the forcing $\bf f$ defined previously) to explain the method. Let's say $\mathbf f$ is the vector with the values of $f(z)$ at the grid points in the non-periodic interval $[0,L_z]$. Then, in order to generate a periodic continuation $\mathbf f^c$ from the original $\mathbf f$ values, one needs to specify two parameters, which are the number of continuation points desired $C$ (i.e., the number of grid points in the extension of the original domain), and the degree of the polynomial basis $d$ (i.e., the number of values of $\mathbf f$ near each boundary to use in the polynomial adjustment, as illustrated in \cref{fc:fig:fc_demo}). Given those parameters, a $C \times d$ continuation matrix $A$ and a $d \times d$ projection matrix $Q$ can be computed (see definitions of these matrices in \cite{Amlani2016}, and the resulting continuation points are obtained with the expression
\begin{equation}
\label{fc:eq:continuation}
\mathbf f^c = A Q \mathbf f_f + A^\ddagger Q^{\varPi} \mathbf f_l,
\end{equation}
where $\mathbf f_f$ (resp., $\mathbf f_l$) are the first (resp., last) $d$ points of $\mathbf f$. In this equation, $^\ddagger$ and $^{\varPi}$ denote the row-reversing and column-reversing operations, respectively. The resulting data $\mathbf f^c$ and its spatial derivatives are periodic in the $[0,L_z')$ interval, and thus can be Fourier transformed efficiently. Applying \cref{fc:eq:continuation} to each $s$-line of $\phi_{nml}$ provides $\phi^c_{nml}$.

It should be noted that both $A$ and $Q$ are independent of the original data $\mathbf f$ (or $\phi$), so they can be precomputed only once before starting the iteration of the Navier-Stokes equations and utilized henceforth. Precomputing these tables is extremely fast, demanding only a couple of minutes in a modern CPU core. The computational cost of calculating the continuations is hence reduced to $2\times N_x \times N_y$ matrix-vector multiplications with very low dimensionality. Even more, these products can be computed in parallel, and each one fits in the L1 cache of a CPU core. For example, in a case with $N_z=991$ we obtain excelent results (both in the order of the approximation as well as in the computational cost) employing $C=33$ and $d=7$.

As previously mentioned, the one-dimensional FC-Gram method summarized above can be applied to the 3D case by utilizing \cref{fc:eq:continuation} for each line with fixed $(x,y)$ coordinates and results, for instance, in a velocity field ``continued" in the $z$ direction $\bm v^c$, and whose Fourier coefficients $\hat{\bm v}^c_{nml}$ can then be obtained via standard 3D-FFT computations with great accuracy. It is straightforward to see that the FFT and the FC-Gram operators commute and hence $\hat{\bm v}^c$ can alternatively be obtained by first transforming $\bm v$ to the mixed $(k_x,k_y,z)$ domain (a 2D-FFT operation) and then performing one dimensional FC-Gram continuations for each $(k_x,k_y)$ coordinate, followed by an additional 1D-FFT calculation. When computing the 3D-FFT in parallel, this latter property is useful to ensure that the continuations are computed locally, severely boosting performance (see more details in Sec.~\ref{sec:numerics}). Further information on the technical details of the FC-Gram transformation method, and on the mathematical properties of the Gram basis of polynomials, can be found in \ref{app:FC-Gram}.
\section{Method for solving the incompressible Navier-Stokes equations}
\label{sec:method}

\subsection{Time-splitting forward Euler method and boundary conditions}

Numerically, one common practice for dealing with the problem of the boundary conditions for the pressure mentioned in \cref{sec:equations} is to employ a time-splitting scheme \cite{Chorin1968, Canuto1988}. The idea behind this technique is to introduce an auxiliary field $\bm v^*$ that verifies the pressureless momentum equation and which, when subtracted the pressure gradient, results in a solenoidal velocity field at the next time step. Although for the simulations in Sections \ref{sec:chan} and \ref{sec:Rayleigh} we employ a higher-order explicit Runge-Kutta integrator, the algorithm for evolving in time the incompressible Navier-Stokes equations will be described here first using a forward Euler method for notational clarity. The generalization to arbitrary order Runge-Kutta will be deferred until the end of this section.

In the time-splitting forward Euler time stepping method, given the velocity field $\bm v^t$ at the time $t$, the velocity at the next time step $\bm v^{t+\Delta t}$ is obtained as
\begin{align}
	\label{met:solver:ts-momentum}
	{\bm v}^{* t+\Delta t} &= \bm v^t + \Delta t \left[\left(\bm v^t \cdot \bm \nabla \right) \bm v^t + \nu \nabla^2 \bm v^{t} + \bm f^{t} \right],\\
	\label{met:solver:ts-poisson}
	\nabla^2 p^{t+\Delta t} &= \frac{1}{\Delta t}\bm \nabla \cdot {\bm v^*}^{t+\Delta t},\\
	\label{met:solver:ts-projection}
	\bm v^{t+\Delta t} &= {\bm v}^{* t+\Delta t} - \Delta t \bm \nabla p^{t + \Delta t}.
\end{align}
This method is similar to methods often used to evolve the Navier-Stokes equations in 3D-periodic domains. However, written in this way, boundary conditions can be imposed separately both to $\bm v^*$ and $p$. This guarantees that after the projection step in \cref{met:solver:ts-projection}, the error of $\bm v^{t+\Delta t}$ at the boundary remains controlled for both the normal and tangential directions.

In particular, different boundary conditions can be supplied to both $\bm v^*$ after Eq.~\eqref{met:solver:ts-momentum}, and to the normal or tangential derivative of $p$ when solving Eq.~\eqref{met:solver:ts-poisson}, depending on the order of approximation desired. For this section we use a modified global $\mathcal{O}(\Delta t^2)$ scheme derived in \cite{Kim1985}, which satisfies the boundary conditions with an error of order $\mathcal{O}(\Delta t)$. We also employ the stability enhancing modification to the wall-normal velocity presented in \cite{Orszag1986}, where the following boundary conditions were introduced
\begin{alignat}{2}
\label{met:solver:bound-v}
{\bm v^*_{\parallel}}^{t+\Delta t} &= \Delta t \bm \nabla_\parallel p^t &\qquad \qquad \text{at }z=0,L_z,\\
\label{met:solver:bound-p}
\dpd{p}{z}^{t+\Delta t} &= \frac{1}{\Delta t} \hat {\bm z} \cdot {\bm v^*}^{t+\Delta t} &\qquad \qquad \text{at }z=0,L_z.
\end{alignat}
Here the symbol $\parallel$ denotes the components tangential to the wall and $\hat {\bm z}$ is the unit vector in the $z$ direction. It should be noted that when coupled with, for example, a second order time integrator (as done in Sections \ref{sec:chan} and \ref{sec:Rayleigh}), these boundary conditions result in a $\mathcal{O}(\Delta t^2)$ slip velocity at the walls, a fact that will be numerically verified.

\subsection{Solution for the pressure in the time-splitting forward Euler method}

Following the method presented in the previous section, for every timestep we start by solving for the pressureless velocity in \cref{met:solver:ts-momentum}. This can be easily accomplished for the continued fields ${\bm v}^{*c}$ in the wavenumber domain, where the evolution equation takes the form
\begin{equation}
\hat{\bm v}^{*c}_{nml} = \hat{\bm v}^c_{nml} + \Delta t \left[ \left[ \reallywidehat{\left(\hat{\bm v}^c \cdot \bm \nabla \right) \hat{\bm v}^c} \right]_{nml} - \nu k_{nml}^2 \hat{\bm v}^c_{nml} + \hat{\bm f}^{c}_{nml} \right].
\end{equation}
As before, hats denote Fourier transformed quantities, and note that time superindices where dropped for clarity. The Fourier coefficients of the non-linear term are obtained by standard pseudospectral calculations in $\mathcal{O}[N_xN_yN_z \log (N_x) \log (N_y) \log (N_z)]$ operations (i.e., by computing the derivatives in Fourier space, the product in real space, and returning to Fourier space) \cite{Orszag1969, Cooley1965}. Computation of derivatives of the continued fields in Fourier space is straightforward, and performed as in pseudospectral methods in periodic domains, by multiplying Fourier coefficients by their corresponding wavenumber and the imaginary unit.

The next step in the algorithm is to transform the velocity field to the $(k_x,k_y,z)$ domain and apply the boundary conditions in \cref{met:solver:bound-v} to the unphysical velocity field $\bm v^*$ via strong imposition (also known as ``injection"). In other words, values of the pressureless velocity at the boundaries are replaced to fulfill \cref{met:solver:bound-v}, after which a FC-Gram is performed, and $\bm v^{*c}$ is transformed back to Fourier space. 

Afterwards, the pressure gradient can be obtained by solving \cref{met:solver:ts-poisson} with boundary condition given by \cref{met:solver:bound-p}. The Poisson equation accepts a homogeneous solution $p^H$ satisfying $\nabla^2 p^H =0$, and an inhomogeneous solution $p^I$ such that $\nabla^2 p^I$ satisfies \cref{met:solver:ts-poisson}. Using periodic boundary conditions in the extended domain, the gradient of the inhomogeneous pressure solution, $\bm \nabla p^{I}$, can be readily computed in the 3D Fourier space as
\begin{equation}
\bm \nabla \hat{p}^{Ic}_{nml} = \frac{\bm k_{nml} \cdot \hat{\bm v}_{nml}^{*c}}{k^2} \bm k_{nml}. 
\end{equation}
This solution does not necessarily satisfy the boundary conditions for the pressure, but $p^H$ can now be used to impose \cref{met:solver:bound-p}. To compute the homogeneous solution $p^H$ we take advantage of the fact that in Cartesian coordinates there exists an analytical solution of $\nabla^2 p^H =0$ in the mixed $(k_x,k_y,z)$ domain with coefficients given by
\begin{equation}
\label{met:solver:pressure-analy}
\hat{p}^H_{nm}(z) = A_{nm} e^{\gamma_{nm} \left(z - L_z\right)} + B_{nm} e^{-\gamma_{nm} z} + D z .
\end{equation}
Here, $\gamma_{nm} = (k_{x,n}^2 + k_{y,m}^2)^{1/2}$, $k_{x,n}=\pi(2n-N_x)/L_x$ and $k_{y,m}=\pi(2m-N_y)/L_y$ as in Eq.~\eqref{eq:k}, and $A_{nm}$, $B_{nm}$ and $D$ are coefficients that depend on the boundary conditions. Differentiating \cref{met:solver:pressure-analy} and using \cref{met:solver:bound-p} it is straightforward to get the following expressions for these coefficients
\begin{align}
\label{met:solver:A}
A_{nm} &= \dfrac{t_{nm}- b_{nm} e^{-\gamma_{nm} L_z}}{\gamma_{nm}\left( 1 - e^{-2\gamma_{nm}L_z}\right)}, \\
\label{met:solver:B}
B_{nm} &= - \dfrac{b_{nm} - t_{nm} e^{-\gamma_{nm} L_z}}{\gamma_{nm}\left( 1 - e^{-2\gamma_{nm}L_z}\right)}, \\
\label{met:solver:D}
D &=  \frac{1}{\Delta t} b_{00} = \frac{1}{\Delta t} t_{00},
\end{align}
where 
\begin{align}
b_{nm} &= \hat {\bm z} \cdot \left. \left[\frac{\hat{\bm v}^*_{nm}(z)}{\Delta t} - \bm \nabla \hat{p}^{Ic}_{nm}(z)\right] \right|_{z=0} , \\
t_{nm} &= \hat {\bm z} \cdot \left. \left[\frac{\hat{\bm v}^*_{nm}(z)}{\Delta t} - \bm \nabla \hat{p}^{Ic}_{nm}(z)\right] \right|_{z=L_z}.
\end{align}
As before, the notation $\hat{\bm v}^*_{nm}(z)$ and $\nabla \hat{p}^{Ic}_{nm}(z)$ denotes quantities are in the mixed $(k_x,k_y,z)$ domain. Also, note that as the boundary conditions are imposed on the derivative of $p$, the total pressure is defined up to a constant that can be safely ignored. An analytical expression for the homogeneous pressure gradient $\bm \nabla p^H$ in the mixed $(k_x,k_y,z)$ domain can then be trivially obtained from \cref{met:solver:pressure-analy} as
\begin{align}
\bm \nabla \hat{p}^H_{nm}(z) =& (i k_{x,n} \bm{\hat x} + i k_{y,m} \bm{\hat y}) \left[ A_{nm}e^{\gamma_{nm} \left(z -L_z\right)} + B_{nm} e^{-\gamma_{nm} z} \right] + \nonumber \\
{}& \gamma_{nm} \left[ A_{nm}e^{\gamma_{nm} \left(z - L_z\right)} - B_{nm} e^{-\gamma_{nm} z} + \frac{D}{\gamma_{nm}} \right] \bm{\hat z}.
\label{met:solver:nablap-homogeneous}
\end{align}
We can FC-Gram continue the above expression and Fourier transform in the $z$ direction to get the coefficients of the ``continued" total pressure gradient $\bm \nabla \hat{p}_{nml}^{c} = \bm \nabla \hat{p}_{nml}^{Ic} + \bm \nabla p_{nml}^{Hc}$, and use \cref{met:solver:ts-projection} to finally obtain the velocity field at the next time step in the $(k_x,k_y,k_z)$ domain.

\subsection{Filtering}

As it is widely known, the computation of non-linear terms by means of pseudospectral calculations generates a pileup of energy in the highest wavenumbers (known as \textit{mode aliasing} in this context), due to the nature of the discrete version of the convolution theorem. For cuadratic non-linearities, a standard way of dealing with this instability is by utilizing the so-called 2/3-rule, in which wavenumbers with $k>2k_\text{max}/3$ are filtered out before and after the pseudospectral calculation \cite{Orszag1971}, and where $k_\text{max}$ is the Nyquist maximum resolved wavenumber. However, in our context, employing such an abrupt filter would result in a serious distortion of the fields in physical space, as high wavenumbers are required to accurately represent the periodically-extended fields. We circumvent this problem by employing an exponential filter of the kind proposed in \cite{Albin2011}, such that for a scalar field $\phi$, its filtered spectral coefficients $\hat \phi_{nml}^f$ become
\begin{equation}
\label{met:eq:filter}
\hat \phi_{nml}^f = \hat \phi_{nml} \exp \left[-\beta \left( \left(\frac{2L_xk_{x,n}}{\pi N_x}\right)^{2p} +  \left(\frac{2L_yk_{y,m}}{\pi N_y}\right)^{2p} + \left( \frac{2L'_z k_{z,l}}{\pi (N_z+C)}\right)^{2p} \right) \right].
\end{equation}
As the largest wavenumber is attenuated by a factor of $e^{-\beta}$, it is natural to choose $\beta=b\ln(10)$, where $b$ is the number of significant decimal digits desired (16 for double precision calculations). The parameter $p$ should be chosen so that the error introduced in the filtering step remains smaller than the one associated to the time-marching scheme. It was shown in \cite{Albin2011} that a value of $2p \ge 55$ suffices to attain an $\mathcal{O}(\Delta t^5)$ approximation for the case of forth order Adams-Bashforth integration. For this work we got good results when choosing the value $2p = 100$, and this is therefore the value used for all our simulations in Sections \ref{sec:chan} and \ref{sec:Rayleigh}.

\subsection{Higher-order time-splitting Runge-Kutta method}
\label{sec:high-order}

With the above discussion in mind it is possible now to introduce a full algorithm for a time-splitting $o$-th order Runge-Kutta integrator, which approximates the no-slip boundary condition with an accuracy $\mathcal{O}(\Delta t^o)$. The method we present also has the advantage of requiring low storage, with only a few arrays stored in computer memory in each iteration:

\vskip .5cm
\noindent \textbf{FOR} $j$ in $1, \hdots, o$:
\begin{enumerate}
\item Evolve the pressureless momentum equation in $(k_x,k_y,k_z)$ space
\begin{equation}
\hat{\bm v}_{nml}^{* t+j\Delta t/o} = \hat{\bm v}_{nml}^t + \frac{j\Delta t}{o} \left[ \left[ \reallywidehat{\left(\bm v^{t+(j-1)\Delta t/o} \cdot \bm \nabla \right) \bm v^{t+(j-1)\Delta t/o}}\right]_{nml} - \nu k_{nml}^2 \hat{\bm v} _{nml}^{t+(j-1)\Delta t/o} \right] ,
\end{equation}
where the continuation superindices $^c$ were dropped for clarity. Note that the filter in \cref{met:eq:filter} must be applied to the non-linear term to prevent aliasing instabilities.
\item Transform $\hat{\bm v}^{*t+j\Delta t/o}$ to the $(k_x,k_y,z)$ domain, and impose
\begin{equation}
{\bm v^*_{\parallel}}^{t+j\Delta t/o} = \frac{j\Delta t}{o} \bm \nabla_\parallel p^{t+(j-1)\Delta t/o}  \qquad \text{at }z=0,L_z .
\end{equation}
\item FC-Gram transform ${\bm v}^{*t+j\Delta t/o}$ back to the $(k_x,k_y,k_z)$ domain, and obtain the inhomogeneous solution for the pressure
\begin{equation}
\bm \nabla \hat{p}^{It+j\Delta t/o}_{nml} = \frac{\bm k_{nml} \cdot \hat{\bm v}^{*t+j\Delta t/o}_{nml}}{k^2} \bm k_{nml}. 
\end{equation}
\item Transform $\hat{\bm v}^{*t+j\Delta t/o} - \bm \nabla \hat{p}^{It+j\Delta t/o}_{nml}$ to the $(k_x,k_y,z)$ domain, and use Eqs.~\eqref{met:solver:pressure-analy} to \eqref{met:solver:B} to solve for the homogeneous solution for the pressure, now with $D=o b_{00}/(j\Delta t) = o t_{00}/(j\Delta t)$, and with the coefficients stemming from the Neumann boundary conditions now taking the form
\begin{equation}
b_{nm} = \bm{\hat z} \cdot \left. \left[ \frac{o}{j\Delta t} \hat{\bm v}^*_{nm}(z) - \bm \nabla \hat{p}^{I}_{nm}(z) \right] \right|_{z=0} , \qquad
t_{nm} = \bm{\hat z} \cdot \left. \left[ \frac{o}{j\Delta t} \hat{\bm v}^*_{nm}(z) - \bm \nabla \hat{p}^{I}_{nm}(z) \right] \right|_{z=L_z} .
\end{equation}
\item FC-Gram transform the homogeneous pressure gradient to the $(k_x, k_y, k_z)$ domain, and project the velocity to the space of solenoidal functions 
\begin{equation}
\bm v^{t+j\Delta t/o} = {\bm v^*}^{t+j\Delta t/o} - \frac{j \Delta t}{o} \bm \nabla p^{t+j\Delta t/o}.
\end{equation}
\end{enumerate}
\noindent end \textbf{FOR}.

\section{Numerical implementation}
\label{sec:numerics}

\begin{figure}[t]
	\centering
	\includegraphics[width=\textwidth, keepaspectratio=true]{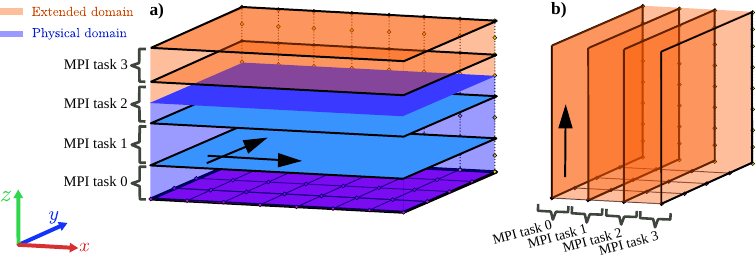}
	\caption{Slab decomposition used for the parallelization. Notation is the same as in Fig.~\ref{met:fig:domain}. (a) The physical domain, with the extended domain in real space. Each MPI task gets one of the slabs (for convenience, only four tasks are shown). Note FFTs in the $x$ and $y$ directions (indicated by the black arrows) can be computed locally with this decomposition. (b) Arrays in Fourier space $(k_x,k_y,k_z)$ or in mixed space $(k_x,k_y,z)$. After the transposition, with this slab decomposition FFTs in the $z$ direction are local, and can be computed without communication. Note the full array in $k_x$ has half the size of the physical domain in $x$, as Fourier transformed arrays are complex while physical data is real.}
	\label{met:fig:slabs}
\end{figure}

The ideas presented in the previous section can be easily implemented in existing parallel pseudospectral codes. For this study we developed a software named Spectral PEriodic Continuation Turbulence SolvER (\textsc{Specter}), freely available at \url{http://github.com/mfontanaar/SPECTER}. \textsc{Specter} is a hybrid OpenMP-MPI-CUDA parallel code written in Fortran 95/2003 (with bindings in C and CUDA), built atop the structure of the Geophyisical High-Order Suite for Turbulence (\textsc{Ghost}) whose parallelization strategy has been shown to present near optimal scaling for over 200.000 CPU cores and 15.000 GPUs \cite{Mininni2011} (see also \url{http://github.com/pmininni/GHOST/}).

The algorithm presented in section \ref{sec:high-order} is easy to implement using a serial FFT. In practice, we use FFTW \cite{Frigo2005} or cuFFT \cite{Nvidia2019} to compute 1D Fourier transforms. FC-Gram continuation operations are encapsulated in the Fourier transform subroutines that call FFTW or cuFFT. Once a pressure solver and FC-Gram routines are implemented, the rest of the code is remarkably similar to any other pseudospectral code commonly used to study isotropic and homogeneous turbulence. As a result, two advantages of the method follow: (1) Obtaining solutions to the evolution of turbulent flows with boundaries in cuboid domains becomes relatively inexpensive (compared with methods for flows in fully periodic domains), as the overhead of computing FC-Gram continuations and of finding a homogeneous solution for the pressure is relatively small. (2) Efficient parallelization of the method is straightforward, as parallelization methods developed for pseudospectral periodic codes can be easily extended to our method.

In particular, \textsc{Specter} uses a 1D domain decomposition (also called a ``slab" decomposition) for the parallelization \cite{Mininni2011}. The domain in real space is partitioned in the $z$ direction, while the Fourier space is partitioned in the $k_x$ direction (see Fig.~\ref{met:fig:slabs}). Each MPI task operates locally over one of these slabs, and MPI is used for operations that require cross-task communication. In particular, most of the communication in the code is done when the 3D-FFTs are computed: in the 1D domain decomposition FFTs of real arrays in the $x$ and $y$ direction are ``local" (i.e., each MPI task has all the information to compute the FFTs in these directions using a serial FFT library), while FFTs in the $z$ direction are non-local. To solve this problem, parallel pseudospectral codes perform a transposition of the data: data is re-arranged in such a way that arrays in Fourier space, both in the $(k_x,k_y,k_z)$ and in the mixed $(k_x,k_y,z)$ spaces, are partitioned in the $k_x$ direction and thus the $z$ direction becomes ``local" for the FFTs (i.e., each MPI task has all the information required to compute the FFTs in this direction in their portion of the domain independently of the other tasks). The transposition involves an all-to-all communication, and is handled in \textsc{Specter} using the same techniques as in \textsc{Ghost} (see \cite{Mininni2011} for more details).

To increase the number of processors that can be used for the parallelization, operations in each MPI task can be further parallelized using OpenMP. To this end, FFTs in each MPI task are performed using multiple threads with FFTW, and all other loops are parallelized using OpenMP pragmas. Moreover, if instead of FFTW, cuFFT is used, FFTs can be computed in CUDA-enabled graphical processing units (GPUs). As this multi-level hybrid parallelization scheme has been described elsewhere, and requires no modification to the Forier Continuation method presented here, we refer the reader to \cite{Mininni2011} for a detailed description as well as for scaling studies. It suffices to say that tests of \textsc{Specter} in parallel environments show no degradation in the parallelization efficiency when compared with the results in \cite{Mininni2011}, and that as a result we expect our method to scale well as the number of processors is increased.

\section{Plane Poiseuille flow simulations}
\label{sec:chan}

\begin{table}
\begin{center}
	\caption{Parameters of the channel flow simulations. ``Run" labels each simulation, $N_x \times N_y \times N_z$ gives the linear resolution in each direction, and $C_z$ and $d_z$ are the number of continuation and boundary matching points in the $z$ direction, respectively. $\overline{\partial_x p}$ is the mean pressure gradient in the $x$ direction, $\nu$ the kinematic viscosity, $\Re$ is the Reynolds number, $\Re_c$ is the centerline Reynolds number, and $\Re_\tau$ is the Reynolds number based on the friction velocity $u_\tau$.}
	\label{chan:tbl:summary}
\begin{tabular}{c c c c S[table-format=-1.1e-1] S[table-format=-1.2e-1] c c c}
\toprule
\toprule
Run & $N_x\times N_y\times N_z$& $C_z$ & $d_z$ & $\overline{\partial_x p}$& $\nu$&  $\Re$& $\Re_c$& $\Re_\tau$\\
\midrule
L1& $32\times 16\times 39$     & 25 & 5 & 2.5e-2 & 2.5e-3  &  330&  260&  23 \\
L2& $128\times 64\times 231$   & 25 & 5 & 1e-2   & 1e-3    &  790&  580&  35 \\
L3& $256\times 128\times 479$  & 33 & 7 & 5e-3   & 5e-4    & 1490& 1010&  52 \\
T1& $256\times 128\times 479$  & 33 & 7 & 2.5e-3 & 2.5e-4  & 2120& 1310&  77 \\
T2& $384\times 192\times 735$  & 33 & 7 & 2.5e-3 & 1.5e-4  & 3480& 2050& 118 \\
T3& $512\times 256\times 991$  & 33 & 7 & 2.5e-3 & 1.25e-4 & 4270& 2500& 146 \\
T4& $768\times 384\times 1503$ & 33 & 7 & 2.5e-3 & 1e-4    & 5380& 3125& 173 \\
\bottomrule
\bottomrule
\end{tabular}
\end{center}
\end{table}

As a validation of the method presented above, we now study the flow between two parallel planes driven by a homogeneous pressure difference along the $x$ (streamwise) direction. The domain is the same as in \cref{met:fig:domain}. This set up is commonly known as plane Poiseuille flow, or simply, channel flow. It is a traditional problem in wall-bounded turbulence, and has been extensively researched both numerically \cite{Kim1987, Moser1999, Hoyas2006, Bernardini2014, Lozano-Duran2014, Lee2015} and experimentally \cite{Eckelmann1974, Lemoult2012, Schultz2013, Klotz2017}. From the numerical point of view, as the flow develops a boundary layer near the walls, it may be argued that a method that refines the grid near the walls (as, e.g., Chebyshev-based pseudospectral methods) can be better suited to study this problem. However, our main aim in this section is to validate the numerical method by comparing with previous results, and thus we consider channel flow as a paradigmatic example. To properly resolve the boundary layer with a uniform grid we will use larger spatial resolutions in the $z$ direction than in $x$ and $y$; note also that as the method we present here can deal in principle with any boundary condition, a refined grid near the walls could be implemented, matching the solutions using FC-Gram transforms between the grid in the center region of the channel and the grid in the near-wall regions. Finally, in spite of the limitations in the resolution, the regular grid allows us to use explicit time stepping with a mild CFL condition, thus partially compensating for extra computational costs.

Imposing a pressure difference between two ends of the box results in a non-periodic pressure inside the domain in the $x$ direction. However, the pressure gradient remains periodic, and can be decomposed as $\bm \nabla p = \bm \nabla p' + \overline{\partial_x p} \bm{\hat x}$, where $\overline{\partial_x p}$ is the mean pressure gradient in the $x$ direction, $\overline{\partial_x p} = (p|_{x=L_x} - p|_{x=0})/L_x$. The effect of the pressure difference at the sides of the box is hence equivalent to a constant forcing in the $x$ direction, and the resulting equations are
\begin{align}
\label{chan:eq:navier-stokes}
\dpd{\bm v}{t} + ({\bm v} \cdot {\bm \nabla}) {\bm v} &= - \bm \nabla p' + \nu \nabla^2 {\bm v} - \overline{\partial_x{p}} \, \bm{\hat x},\\
\label{chan:eq:incompressibility}
\bm \nabla \cdot \bm v &= 0.
\end{align}
This system admits an analytical laminar solution in the limit in which the non-linearities are negligible, given by 
\begin{equation}
\label{chan:eq:laminar}
\bm v=-\frac{\overline{\partial_x p}}{2\nu}z(z-L_z) \bm{\hat x}.
\end{equation}
However, as viscosity decreases and non-linear effects become more relevant, the flow is no longer laminar and turbulence develops. To quantify the strength of non-linear effects, the Reynolds number $\Re$ is commonly employed. This dimensionless parameter is defined as the ratio between non-linear and diffusive terms
\begin{equation}
    \Re = \frac{L_z\overline{v_x}}{\nu} ,
\end{equation}
where $\overline{v_x}$ is the vertically averaged streamwise speed. Alternatively, one can define a centerline Reynolds number as
\begin{equation}
    \Re_c = \frac{\delta \, v_x|_{z=\delta}}{\nu} ,
\end{equation}
where $\delta=L_z/2$ is the box half-height. Although the exact value of the critical Reynolds number $\Re_\text{crit}$ for which the system becomes turbulent depends on the way in which the laminar solution is perturbed (as well as on the precise definition of the Reynolds), typical values for $\Re_\text{crit}$ are in the range $\approx 1700$ to 2300 \cite{Pope2000}.

\begin{figure}
	\centering
	\includegraphics[width=.6\textwidth,keepaspectratio=true]{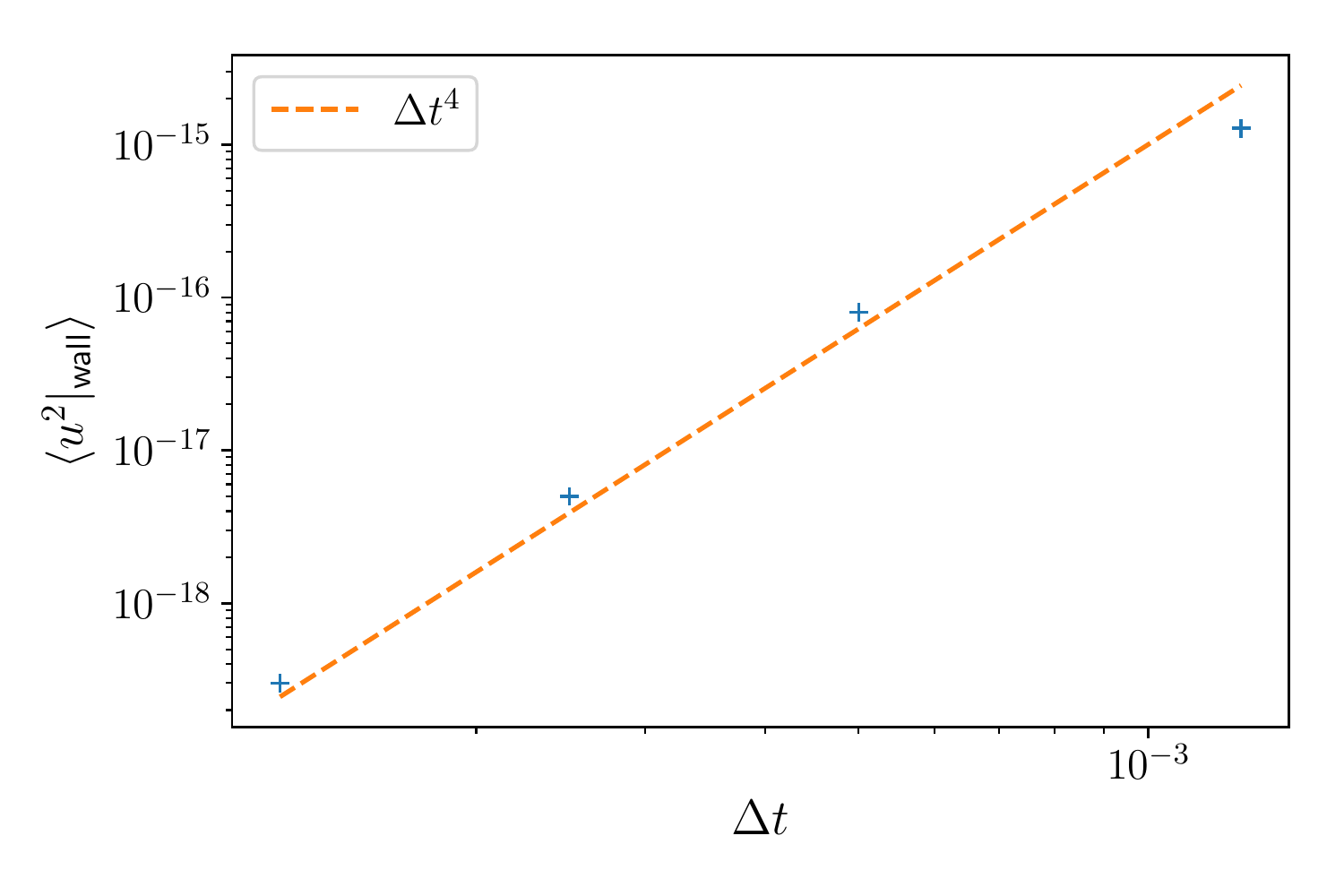}
	\caption{Mean squared streamwise velocity at the wall for simulation T1 as a function of the timestep $\Delta t$ (blue marks). The scaling $\Delta t^4$ is shown as a reference by the dashed line. The point to the right with the largest timestep has a CFL condition of $0.625$.}
	\label{chan:fig:convergence}
\end{figure}

\begin{figure}
	\centering
	\includegraphics[width=.9\textwidth,keepaspectratio=true]{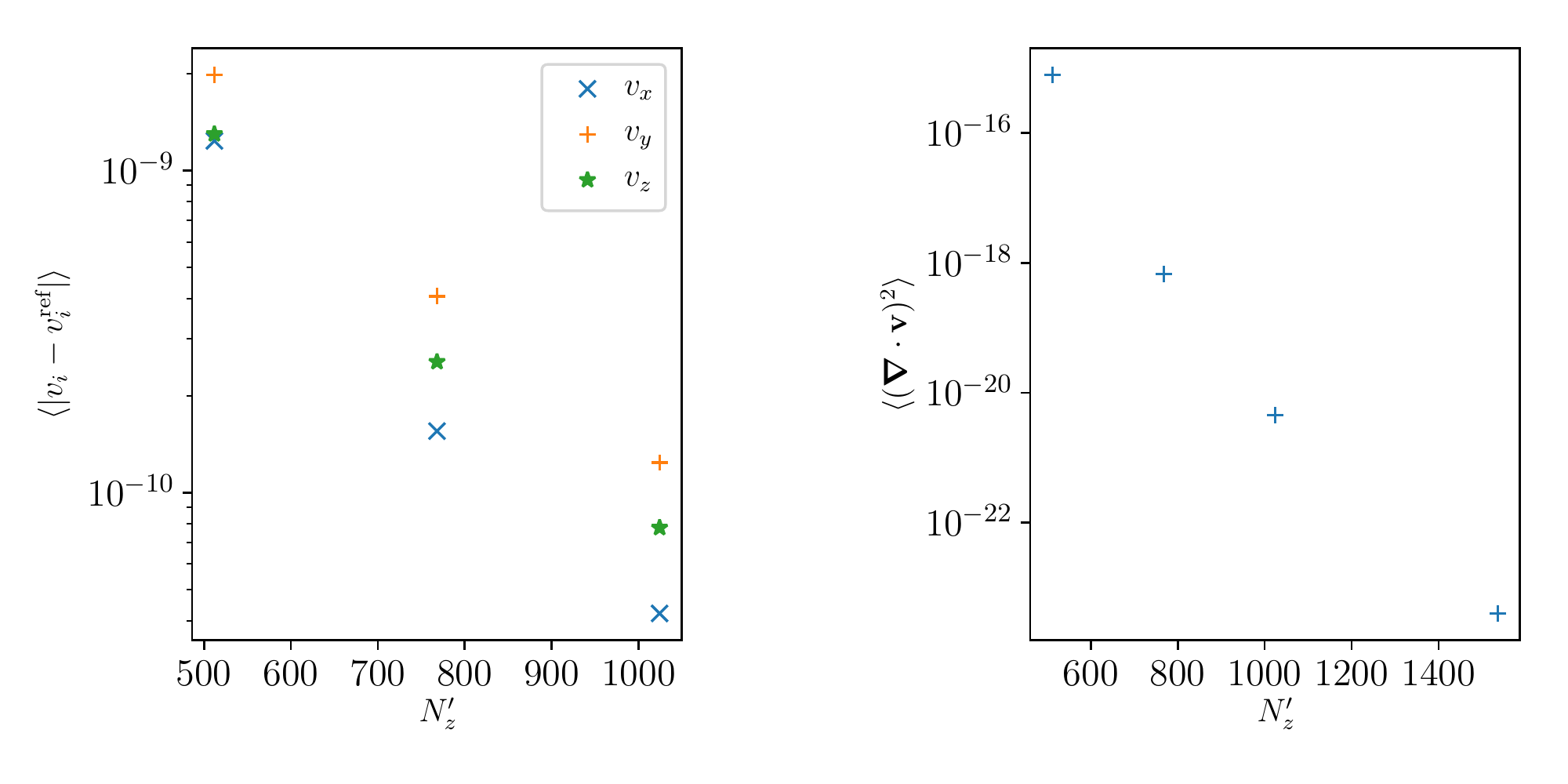}
	\caption{Left: Mean $L_1$ error for each velocity field component as a function of the number of grid points $N'_z$ (counting continuation points) in the vertical non-periodic direction, computed after evolving the steady-state turbulent solution T1 for one large-scale turnover time. All simulations have $C_z=33$ continuation points, so the actual vertical resolution is $N_z=N_z'-33$. A simulation with $N_z'=1536$ is considered as the reference velocity field $\bm{v}^\text{ref}$. The same time step and horizontal resolution is employed for all the simulations. Right: Mean squared divergence of the velocity field for the same set of simulations as a function of the number of vertical grid points $N'_z$.}
	\label{chan:fig:space_convergence}
\end{figure}

\begin{figure}
	\centering
	\includegraphics[width=\textwidth,keepaspectratio=true]{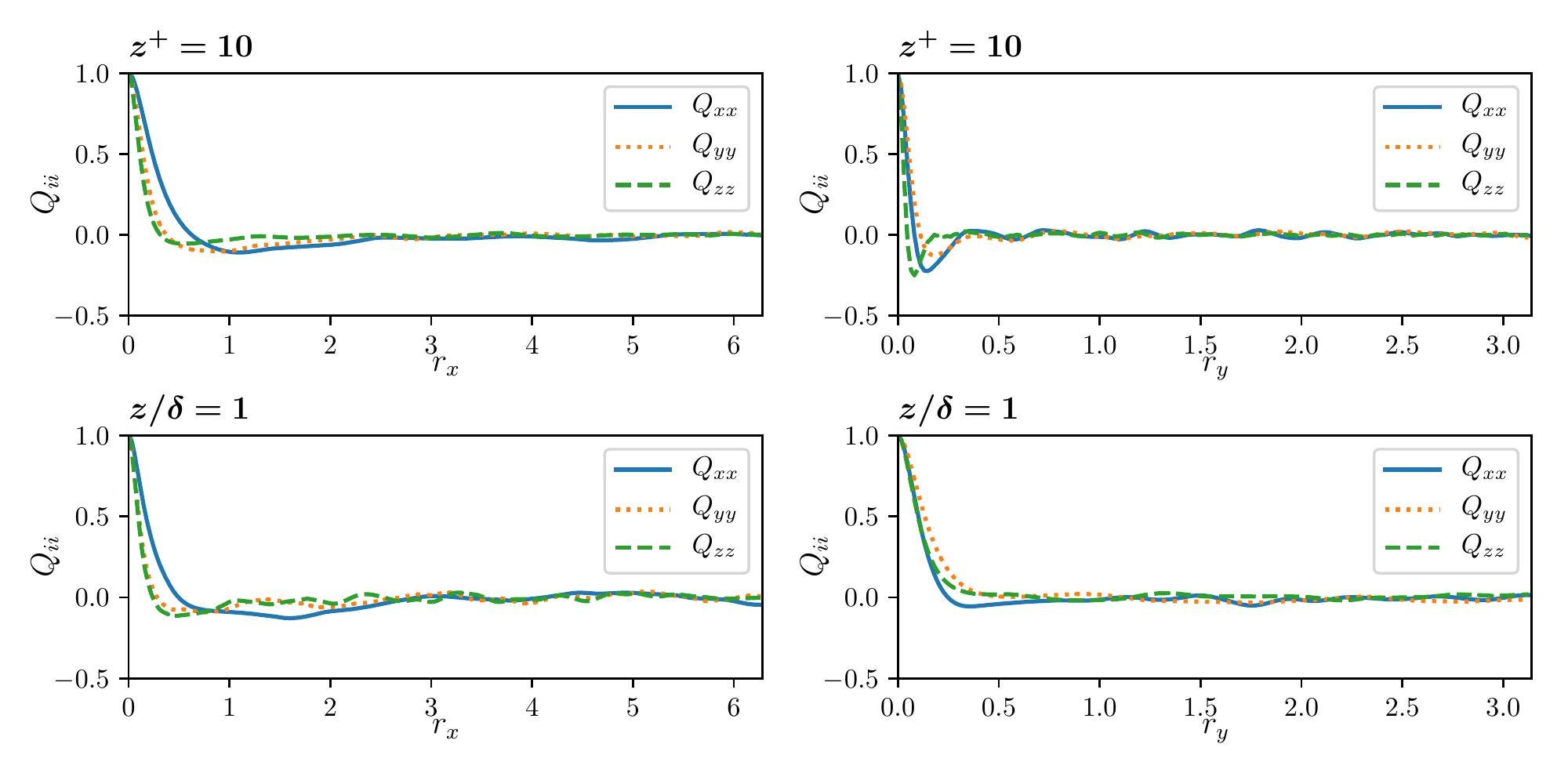}
	\caption{Diagonal elements of the two-point correlation tensor for run T4 as a function of streamwise (left) and spanwise (right) separations, near the wall at 10 wall units (top), and at the centerline (bottom).}
	\label{chan:fig:two_point}
\end{figure}

Using the \textsc{Specter} code we solved \cref{chan:eq:navier-stokes,chan:eq:incompressibility} inside a box of size $L_x \times L_y \times L_z = 4\pi \times 2\pi \times 1$, with $z$ the only non-periodic direction. A total of seven direct numerical simulations were performed, for Reynolds numbers $\Re$ ranging from 300 to 5500, resulting in both laminar and turbulent solutions. A summary of the parameters and the resolution used in each simulation is presented in \cref{chan:tbl:summary}. For all the laminar simulations (L1 to L3) the ratio $\overline{\partial_x p}/\nu$ was maintained equal to 10, in order for the maximum velocity to be of order 1, whereas for the turbulent runs (T1 to T4) the mean pressure gradient $\overline{\partial_x p}$ was kept constant across runs, leaving the centerline velocity as a free parameter. All simulations were integrated in time using a second order version of the method presented in \cref{sec:high-order}.

\begin{figure}
	\centering
	\includegraphics[width=.9\textwidth,keepaspectratio=true]{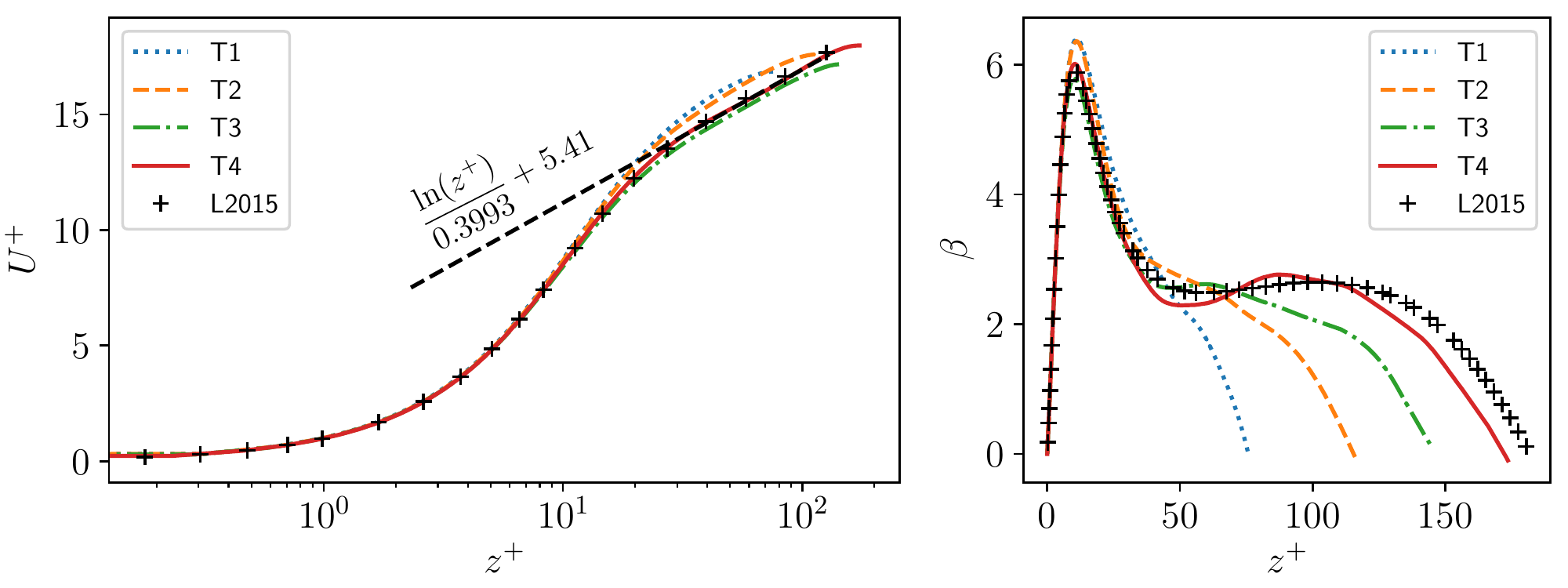}
	\caption{Left: Mean vertical profile in wall coordinates of the streamwise velocity for turbulent runs T1 to T4. A logarithmic profile is shown as a reference by the dashed line. Right: Indicator function for runs T1 to T4. In both cases data for a simulation from L2015 is shown for comparison, and marked with black crosses.}
	\label{chan:fig:profile}
\end{figure}

The first laminar simulation L1 was started from a fluid at rest, and was evolved in time until a steady state was reached. We verified the agreement of the velocity profile with the analytical solution in \cref{chan:eq:laminar}. The last output of this simulation was then scaled up to the resolution of run L2 using spectral interpolation, by zero padding the transformed fields in the wavenumber domain. To the resulting velocity field, a random perturbation with 10\% amplitude was added at the largest scales, and used as initial condition for the higher Reynolds simulation L2. Generation of a random perturbation that is solenoidal and satisfies the boundary conditions requires some care; in the \ref{app:noise} we describe a method that allows generation of noise under these conditions. This procedure was subsequently repeated to generate the initial conditions for run L3 from the steady state of run L2, and for run T1 from the last output of run L3. For all laminar simulations (L1 to L3), a steady parabolic profile was obtained at late times for the streamwise velocity $v_x$ (not shown). For the case of simulation T1, after evolving the system for more than 100 large-scale turnover times, it was clear that a convergence towards a parabolic profile was no longer observed, and a turbulent-like regime was obtained instead. The transition at $\Re\approx2000$ is compatible with the experimental and numerical studies mentioned above. From the last output of run T1 we started (after rescaling) run T2, without a perturbation as the flow was not laminar anymore, and the same procedure was used to start the following runs.

For the first turbulent simulation T1 we also performed time evolution of the flow employing four different timesteps, namely $\Delta t = 1.25\times10^{-3}$, $5\times10^{-4}$, $2.5\times10^{-4}$, and $1.25\times10^{-4}$, in order to study the dependence of the error in the slip velocity. As a reference, note the CFL condition requires $\Delta t \le 2\times10^{-3}$. The result of this analysis is presented in \cref{chan:fig:convergence} where the mean squared streamwise velocity at the wall $\expval{u^2|_\text{wall}}$ is shown as a function of the timestep, along with the predicted scaling of $\Delta t^{-4}$ (corresponding to the expected $\Delta t^{-2}$ scaling for the r.m.s.~error). A good level of agreement with the theoretical prediction is found. Performing the same analysis for the spanwise wall velocity leads to the same results. On the other hand, the error in the mean squared normal velocity at the wall $\expval{w^2|_\text{wall}}|$ was found to be $\mathcal{O}(10^{-35})$ and, as expected, independent of $\Delta t$. This behavior was also observed in the mean squared divergence $\expval{(\bm \nabla \cdot \bm v)^2}$, for which a timestep-independent value of $\mathcal{O}(10^{-16})$ was obtained in all these simulations at fixed spatial resolution.

Also for simulation T1, we studied the spatial convergence of the method as the vertical resolution was increased, by comparing results with increasing resolution against a high resolution numerical solution with the same parameters. To this purpose, we performed simulations with a varying number of vertical grid points in the extended domain, namely $N'_z = 512$, $768$, $1024$, and $1536$, while fixing the number of horizontal grid points to $N_x \times N_y = 256 \times 128$, the matching points to $d_z=7$, and the continuation points to $C_z=33$ (note these choices result in $N_z = N_z'-33$ vertical grid points in the physical domain). The time step was fixed in all simulations to $\Delta t = 2\times10^{-4}$. All simulations started from the same initial condition, corresponding to an output of simulation T1 at $t=82$, which was converted to the required grid size in each case employing spectral interpolation. After integration for one large-scale turnover time, the final states of the simulations with $N'_z = 512$, $768$, and $1024$ were interpolated spectrally to the grid with $N'_z = 1536$. The spatial average of the $L_1$ pointwise errors $\langle | v_i ({\bf x},t) - v_i^\text{ref}({\bf x},t) | \rangle$ (where $i$ is the velocity field component) was then computed using the simulation with $N'_z = 1536$ as the reference solution $\bm{v}^\text{ref}$. The result is shown in \cref{chan:fig:space_convergence}, together with the mean $L_2$ error in the divergence of the velocity field for all values of $N'_z$. High order convergence is observed. Note that we focused here in errors as a function of the vertical resolution. Convergence as a function of horizontal resolution is not shown, as a traditional Fourier pseudospectral method is used in $x$ and $y$, and convergence of this method is well characterized in, e.g., \cite{Canuto1988}. For details of convergence of the Runge-Kutta method as $\Delta t$ is varied, see also \cite{Canuto1988}.

\subsection{Analysis of the turbulent simulations}
We now discuss the results for the turbulent simulations T1 to T4. To this purpose it is useful to introduce the turbulent velocity field components $v'_i = v_i - V_i$, with $V_i = \expval{v_i}_{x,y,t}$ the $i$-th velocity component averaged over $x$, $y$ and $t$, as well as the standard notation $u=v_x$, $v=v_y$, $w=v_z$ (and the corresponding averages when capitalized). Additionally, besides the classical Reynolds number previously mentioned, turbulent channel flow can be better characterized by the friction Reynolds number
\begin{equation}
\Re_\tau = \frac{u_\tau \delta}{\nu},
\end{equation}
where $u_\tau$ is the friction velocity, $u_\tau = \sqrt{\nu (\partial_z U)|_{z=0}}$. Also using $u_\tau$, the following dimensionless variables can be constructed
\begin{equation}
z^+ = \frac{z \nu}{u_\tau}, \qquad \qquad u^+ = \frac{u}{u_\tau},
\end{equation}
with which different results from experiments and simulations can be more directly compared.

Using the turbulent velocity fields, for each simulation we first compute the diagonal elements of the two-point correlation tensor
\begin{equation}
\label{chan:eq:two_point}
Q_{ij} (z,r_l) = \frac{\expval{v_i'(\bm x) v_j'(\bm x + r_l \bm{\hat r}_{l})}_{x,y,t}}{\expval{v_i'(\bm x) v_j'(\bm x)}_{x,y,t}},
\end{equation}
where $r_l$ is a spatial displacement, and $\bm{\hat r}_{l}$ is the unit vector in the direction of the displacement. In the periodic directions the correlations $Q_{ii}(z,r_l)$ are expected to decay to zero for separations $r_l$ of, at most, half the box length. Indeed this can be observed in \cref{chan:fig:two_point}, where all the diagonal elements of the correlation tensor were plotted as a function of both $r_x$ and $r_y$, and both near the bottom wall (for $z^+=10$) and in the center of the channel (i.e., for $z/\delta = 1$).

\begin{figure}
	\centering
	\includegraphics[width=\textwidth,keepaspectratio=true]{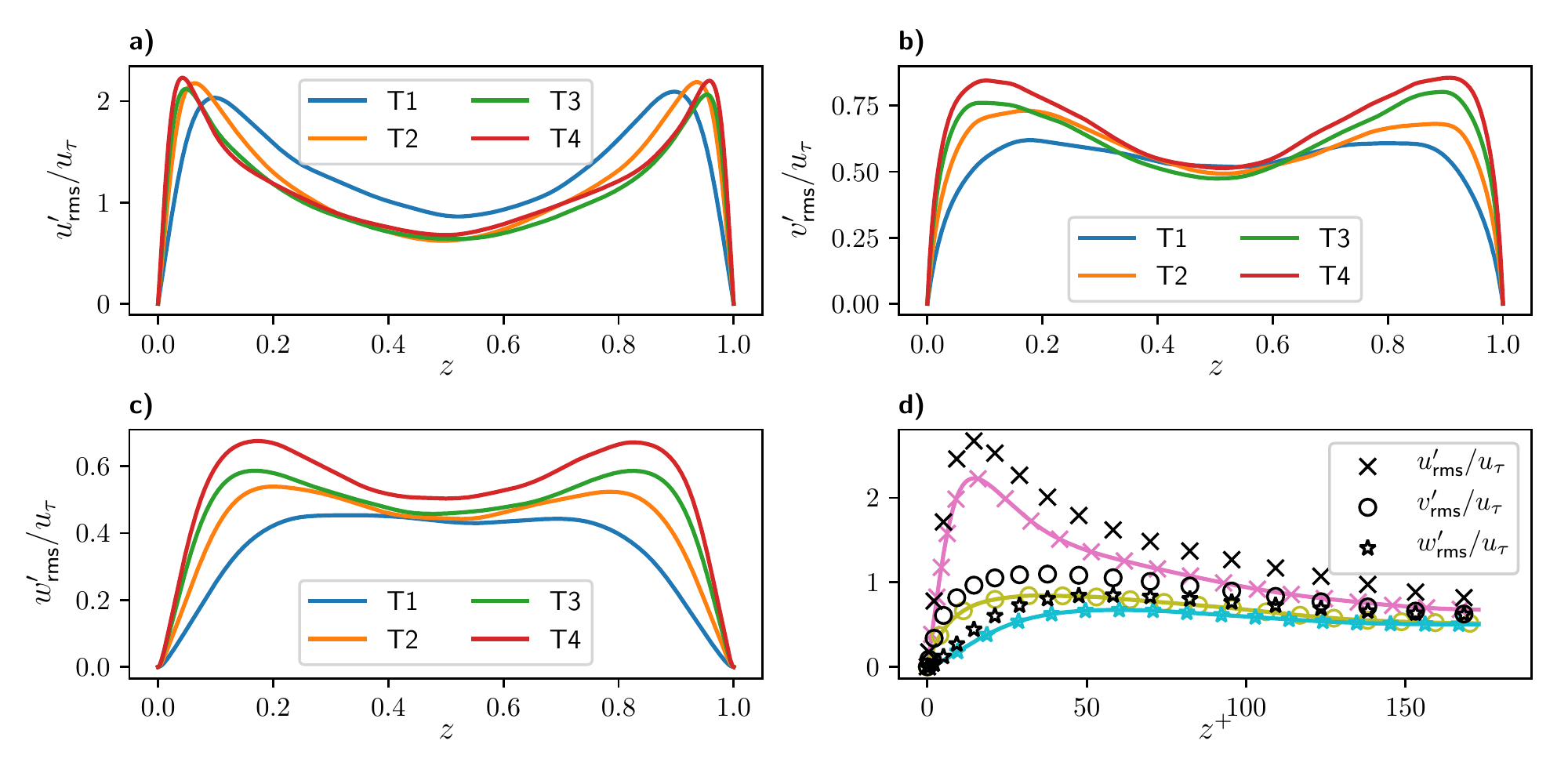}
	\caption{Mean profiles of the root mean squared turbulent velocities normalized by $u_\tau$ and as a function of $z$ for (a) $u'_\textrm{rms}$, (b) $v'_\textrm{rms}$, and (c) $w'_\textrm{rms}$. (d) Comparison (for $z$ in wall units) of the normalized r.m.s.~turbulent velocities between simulation T4 (connected by solid lines) and L2015 (with unconnected symbols).}
	\label{chan:fig:turbulence}
\end{figure}

Expressed in dimensionless units, the mean profile for the streamwise velocity in a fully developed turbulent channel flow has long been proposed to follow a universal logarithmic law far from the walls (known as the von Kármán law), a fact that both experiments and simulations support \cite{Pope2000}. This means that, for sufficiently large $\Re_\tau$, for $z^+\gg 1$ and for $z\ll \delta$, the law
\begin{equation}
U^+ = \frac{1}{\kappa} \ln(z^+) + B ,
\end{equation}
should hold, where $\kappa$ is the von Kármán constant. To verify our simulations are compatible with this law, we estimate the mean profiles for each of our turbulent simulations, shown in \cref{chan:fig:profile}. For the case of T1 and T2 transitional effects are still dominant, and no intervals compatible with a logarithmic law are present. For T3 and T4, on the other hand, the mean profile is in reasonable agreement with the logarithmic law. A best fit of the data for T4 yields $\kappa = 0.3993 \pm 0.0009$ and $B=5.41\pm 0.02$, values which are in agreement with previous studies \cite{Pope2000, LeeSimulation} (the curve corresponding to these values is shown as a reference in \cref{chan:fig:profile}). Moreover, for additional validation of the mean profile, we compare our results with publicly available data from a channel flow simulation at $\Re_\tau = 182$ performed by Lee and Moser \cite{LeeSimulation} (abbreviated L2015 from now on), done at a resolution of $1024\times192\times512$ grid points in a domain of size $8\pi\times2\times3\pi$ using a Fourier-Galerkin method in the periodic directions and a 7\textsuperscript{th} order B-spline collocation in the wall-normal direction, which in their case is along the $y$ axis.  This data is available at \url{http://turbulence.oden.utexas.edu/channel2015/content/Data_2015_0180.html}. 	Their mean profile is also shown in \cref{chan:fig:profile}. Even though for our simulations the aspect ratio of the box is somewhat different and the corresponding values of $Re_\tau$ differ from those in L2015, for the case of run T4 ($\Re_\tau=173$) a considerable interval of agreement in the mean profile is found.

To further analyze the presence of a logarithmic law in our mean profiles, we calculate the so-called indicator function $\beta$ \cite{Lee2015}, defined as
\begin{equation}
\beta(z^+) = z^+ \frac{\partial U^+}{\partial z^+}.
\end{equation}
The indicator function is flat and equal to the von Kármán constant $\kappa$ when a log-law is present. In \cref{chan:fig:profile} a plot of $\beta(z^+)$ is shown for all the simulations. Although, in agreement with our previous observation, no horizontal interval is found for T1 and T2, an approximately flat interval is found for run T3 and an even larger plateau for T4, suggesting that a logarithmic law is indeed compatible with our results. \Cref{chan:fig:profile} also shows a comparison of $\beta$ with the data from L2015 at a larger value of $\Re_\tau$, indicating a good agreement for this quantity near the boundary layer (small values of $z^+$), and the correct trend as $\Re_\tau$ is increased in the log-region.

\begin{figure}
	\centering
	\includegraphics[width=.9\textwidth,keepaspectratio=true]{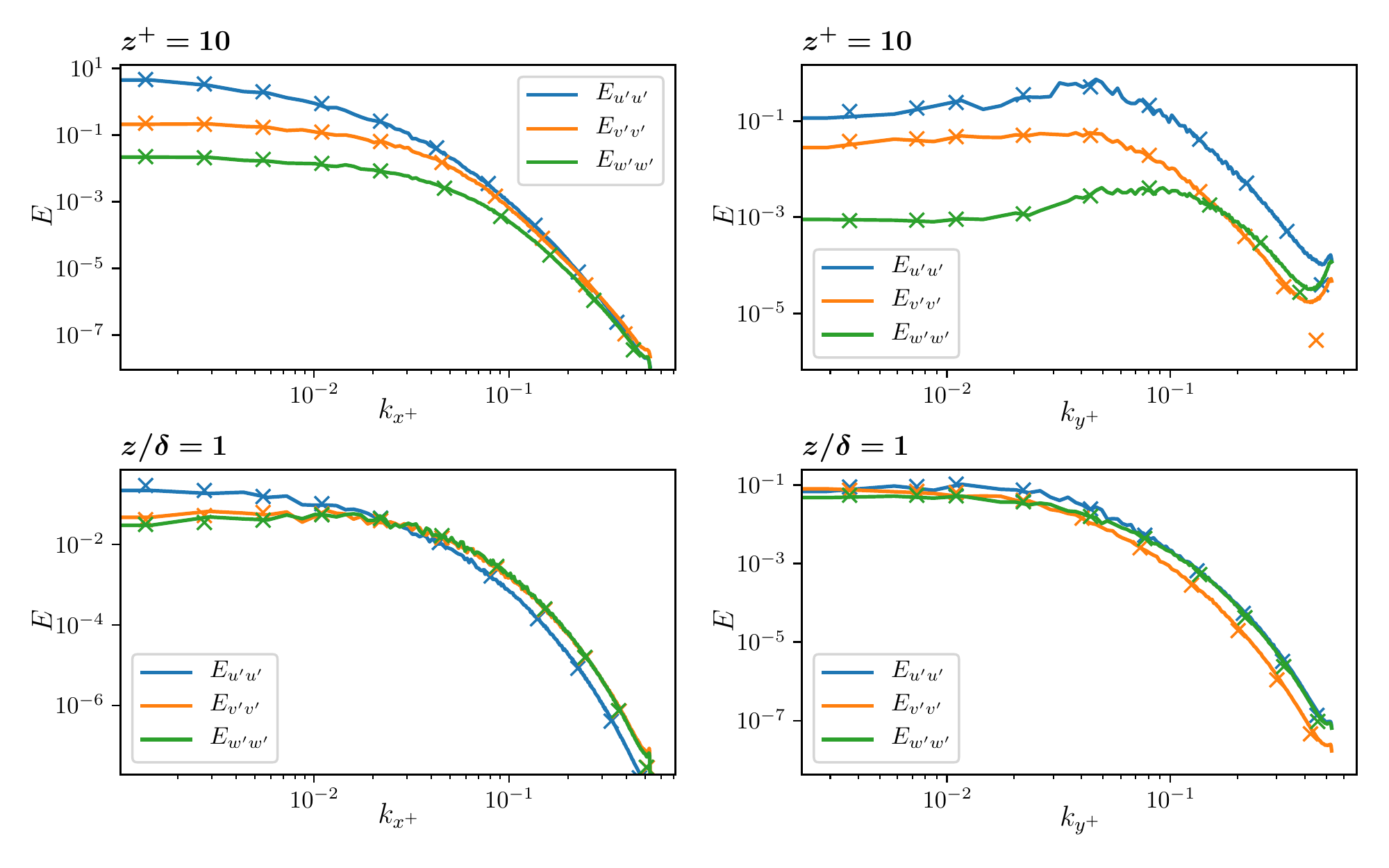}
	\caption{Energy spectra $E_{v_i'v_i'}$ for $u'$, $v'$, and $w'$, as a function of the streamwise (left) and spanwise (right) wavenumbers, both near the wall (top, for $z^+=10$) and at the centerline (bottom, for $z/\delta=1$) for simulation T4 (solid lines). All quantities are expressed in wall units. Spectra from L2015 is marked with the same colors but using unconnected symbols.}
	\label{chan:fig:spectra}
\end{figure}

In \cref{chan:fig:turbulence}, the profiles for the root mean squared turbulence velocity normalized by the friction velocity, $v'_{i,\textrm{rms}}/u_\tau = (\expval{v_i'^2}_{x,y,t})^{1/2}/u_\tau$, are shown. The first thing to notice is that for all turbulent velocity components and for all the simulations, a good degree of symmetry is observed, indicating that an adequate time span was considered for the computation of the mean quantities. Additionally, as expected, the maximum amplitude of the streamwise turbulent fluctuations is located closer to the wall and presents greater values for larger Reynolds numbers. The centerline turbulent fluctuations, nonetheless, seem to decrease for higher $\Re_\tau$ but it soon reaches an approximately constant value. For the spanwise and wall normal components a monotonous increase of the turbulence intensity near the wall is observed conforming $Re_\tau$ grows. When comparing the turbulent intensities for simulation T4 with those observed in L2015 (also in \Cref{chan:fig:turbulence}), a good level of agreement is obtained considering the different values of $\Re_\tau$ between simulations, even though the T4 run was evolved for a shorter period of time and the estimation of the mean amplitude can be affected by finite statistical sampling.

\begin{figure}
	\centering
	\includegraphics[width=\textwidth,keepaspectratio=true]{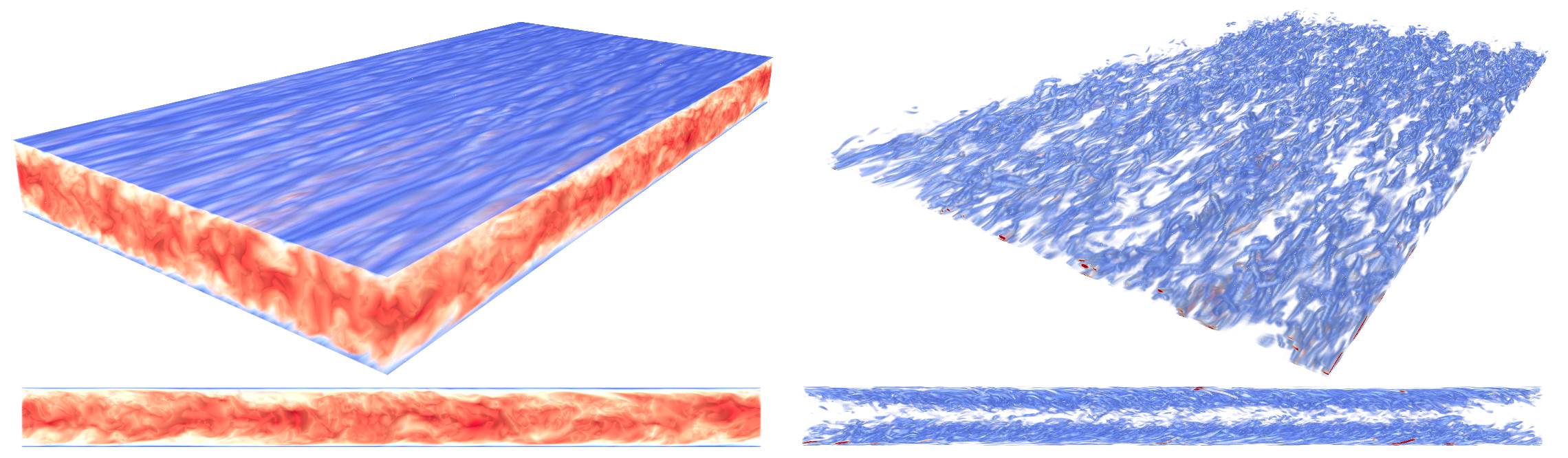}
	\caption{3D renderings of the streamwise velocity in the entire domain (left), and of the r.m.s.~vorticity near the bottom wall (right) in the channel flow simulation T4. The renderings on the bottom show the same quantities from a side view. Blue corresponds to small values, red to high values of each quantity.}
	\label{chan:fig:render}
\end{figure}

We also study the scale dependence of the turbulent fluctuations. To this purpose the 1D energy spectra of the turbulent velocity components (normalized by the friction velocity) is computed. These spectra are defined as
\begin{align}
	\label{chan:eq:spectrax}
	E_{v_i'v_i'} (k_x,z) &= \sum_{k_y} \abs{\hat{v}_i(k_x,k_y,z)}^2, \\
	\label{chan:eq:spectray}
	E_{v_i'v_i'} (k_y,z) &= \sum_{k_x} \abs{\hat{v}_i(k_x,k_y,z)}^2.
\end{align}
To enable a better comparison with the spectra obtained in L2015, which uses a different domain size, we express the energy spectra as a function of the dimensionless wavenumbers $k_i^+ = k_i \nu/u_\tau$. The result for simulation T4 is shown in \Cref{chan:fig:spectra} for two different heights, one near the height of maximum r.m.s. turbulent velocity at $z^+=10$, and the other at the center of the channel (i.e., at $z=\delta$). A smoothly decaying spectra is obtained in all the cases, with the exception of the spanwise spectra near the wall, for which a small accumulation of energy at the largest wavenumbers can be seen, resulting from aliasing. The fast drop that follows in all spectra for larger wavenumbers is associated to our filter. When compared to the spectra from L2015, a significant level of agreement is observed, with very similar decay rates for all velocity components, and for all wavenumbers considered. Even more, some specific features are also reproduced, like the crossing between $E_{u'u'}(k_x^+,0.5)$ and $E_{w'w'}(k_x^+,0.5)$ at $k_x^+ \approx 2\times10^{-2}$, and the accumulation of energy present in $E_{u'u'}(k_y^+)$ and $E_{w'w'}(k_y^+)$ near the wall for intermediate wavenumbers.

Finally, \cref{chan:fig:render} shows 3D renderings of the streamwise velocity and of the r.m.s.~vorticity using the software \textsc{Vapor} \cite{Clyne_2007}. Note the small instantaneous streamwise velocities (indicated by blue) near the boundaries, and the streaks in the streamwise direction. The generation of vorticity near the wall, and the formation of vortex tubes, is more clearly seen in the renderings of the r.m.s.~vorticity, with the largest values of this quantity (indicated by red) taking place near the walls.

Overall, the results shown so far for channel flow simulations indicate our method can give solutions to this problem that are in good agreement with previous studies using pseudospectral or high order B-spline methods, while using a regular grid and (as a result) relatively larger time steps. Solution for the pressure in our method is also relatively straightforward, with a minimum overhead compared with pseudospectral methods in fully periodic domains. However, as already mentioned, channel flow simulations require sufficient resolution near the walls, and as a result we have used a large resolution in the vertical direction to properly resolve the boundary layer. In the next section we consider a problem for which a regular grid provides more numerical advantages.
\section{Plane Rayleigh-Bénard convection simulations}
\label{sec:Rayleigh}

\begin{table}
\begin{center}
	\caption{Parameters of the Rayleigh-Bénard convection simulations. ``Run" labels each simulation, $N_x \times N_y \times N_z$ gives the linear resolution in each direction, and $C_z$ and $d_z$ are the number of continuation and boundary matching points in the $z$ direction, respectively. The kinematic viscosity and thermal diffusivity are $\nu$ and $\kappa$, respectively. Finally, $\gamma=\sqrt{(\alpha g \Delta T/h)}$ (where $\alpha$ is the fluid thermal expansion coefficient, $g$ the acceleration of gravity, $\Delta T$ the temperature difference between the plates, and $h$ the distance between the plates), $\Ra$ is the Rayleigh number, and $\Nu$ is the Nusselt number.}
	\label{rayben:tbl:convection}
\begin{tabular}{{c c c c S[table-format=-1.1e-1] S[table-format=-1.1e-1] c S[table-format=-1.1e-1] S S }}
\toprule
\toprule
Run& $N_x\times N_y\times N_z$& $C_z$ & $d_z$ & $\nu$& $\kappa$& $\gamma$& $\Ra$& $\Nu$ & $\Nu \Ra^{1/3}$\\
\midrule
C1& $ 256 \times  256 \times 103$ & 25 & 5 &   1e-3 &   1e-3 & 1 &   1e6 &  6.55 & 0.071 \\
C2& $ 512 \times  512 \times 231$ & 25 & 5 &   5e-4 &   5e-4 & 1 &   4e6 & 10.89 & 0.069 \\
C3& $1024 \times 1024 \times 479$ & 33 & 7 & 2.5e-4 & 2.5e-4 & 1 & 1.6e7 & 17.99 & 0.066 \\
\bottomrule
\bottomrule
\end{tabular}
\end{center}
\end{table}

\begin{figure}
	\centering
	\includegraphics[width=\textwidth,keepaspectratio=true]{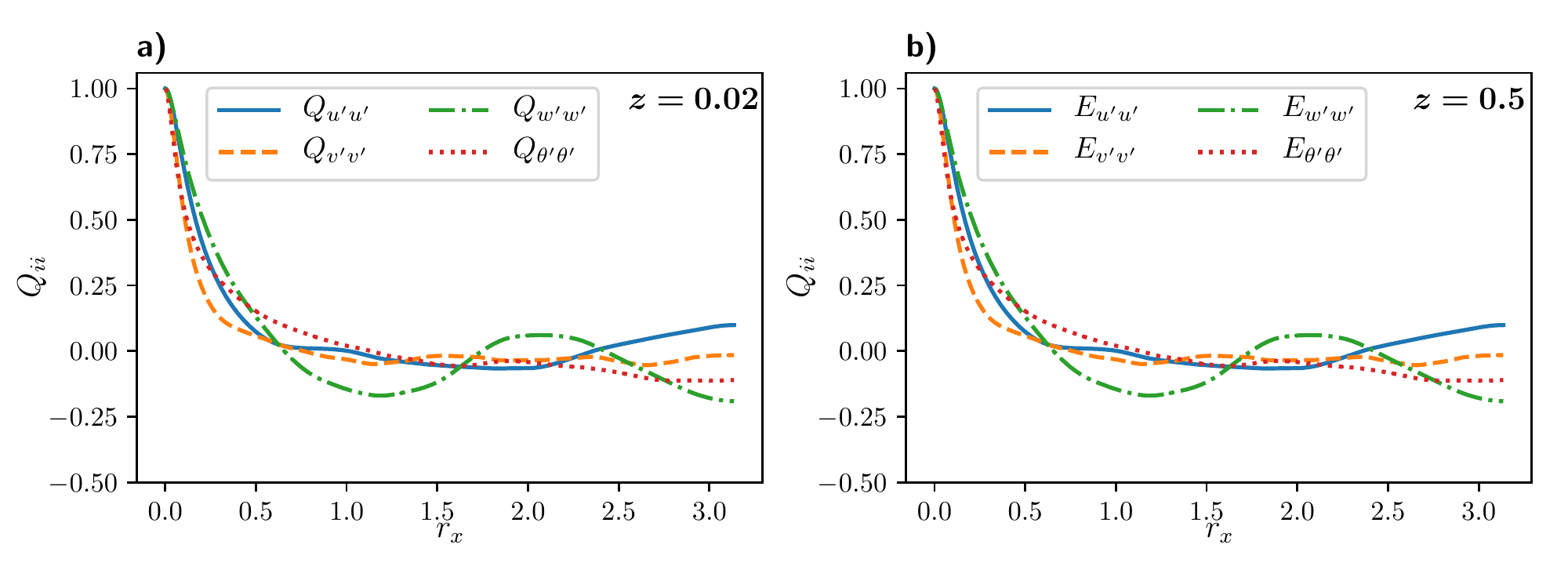}
	\caption{Diagonal elements of the two point correlation tensor $Q_{ii}$ for simulation C3 at \textbf{a)}: the thermal boundary layer $z=\delta_\theta=0.02$ and \textbf{b)} the center of the box $z=0.5$.}
	\label{rayben:fig:rb_two_point}
\end{figure}

We now analyze the application of the method presented above to the case of another set of equations. To this purpose we study the turbulent plane Rayleigh-Bénard convection problem \cite{Biringen1990, Bodenschatz_2000, Amati2005, Lohse_2010,Belmonte1994}, where the space between two plates held at constant temperature is filled with a fluid. If the bottom plate has a fixed temperature $T_b$ much greater than the temperature at the top $T_t$, the fluid destabilizes and convection develops, displaying the well known Bénard cells. Further increasing the temperature difference between the plates leads to turbulent convection \cite{davidson_2013}. For incompressible fluids, it is not unusual to study the case in which density variations with temperature are so small that they only affect the dynamics of the fluid via the buoyancy force, leading the incompressible Bousinessq equations
\begin{align}
    \label{rayben:eq:velocity}
    \dpd{\bm{v}}{t} + (\bm{v} \cdot \bm{\nabla}) \bm{v} &= - \bm \nabla p - \gamma \theta \bm{\hat z} + \nu \nabla^2 \bm{v} ,\\
    \label{rayben:eq:incomp}
    \bm \nabla \cdot \bm v &= 0 ,\\
    \label{rayben:eq:theta}
    \dpd{\theta}{t} + (\bm{v} \cdot \bm{\nabla}) \theta &=  \gamma v_z + \kappa \nabla^2 \theta .
\end{align}
Here, the total temperature is $T = T_0+T'$, where $T_0$ is a linear background profile $T_0(z) = T_b-z\Delta T/h$ with $h=L_z$ the height of the domain, $\Delta T=T_b - T_t$ the temperature difference between the plates (with $T_b$ and $T_t$ the temperatures at the bottom and top plates, respectively), and $T'$ the ``fluctuating" temperature that corrects the background profile. The total pressure is $P=p_0+p$, with $p$ the correction to a background hydrostatic pressure $p_0=\alpha gz (T_b  - z\Delta T/h) + P_0$, with $\alpha$ the fluid thermal expansion coefficient, $g$ the acceleration of gravity, and $P_0$ a constant (as before, the mean mass density of the fluid is $\rho=1$ in dimensionless units). In Eqs.~\eqref{rayben:eq:velocity} and \eqref{rayben:eq:theta} we write the correction to the total temperature in units of velocity by defining $\theta = T' \sqrt{\alpha g h/\Delta T}$. Finally, $\gamma = \sqrt{\alpha g \Delta T/h}$, and $\nu$ and $\kappa$ are respectively the kinematic viscosity and the thermal diffusivity. With these choices the boundary conditions for $\bm v$ and $\theta$ are periodic in the $x$ and $y$ directions, while $\bm v = \bm 0$ and $\theta = 0$ at both $z=0$ and $z=h$.

The system of \cref{rayben:eq:velocity,rayben:eq:incomp,rayben:eq:theta} has an important dimensionless number that controls the instability of the system to natural convection, the Rayleigh number defined as
\begin{equation}
    \textrm{Ra} = \frac{\alpha g h^3 \Delta T}{\nu \kappa} = \frac{\gamma^2 h^4}{\nu \kappa}.
    \label{rayben:eq:rayleigh}
\end{equation}
Other important dimensionless numbers are the Prandtl number, $\textrm{Pr} = \nu/\kappa$, which is 1 in all our simulations, and the Nusselt number which quantifies the ratio of convective to conductive heat transfer, defined in \cite{Belmonte1994} as
\begin{equation}
    \textrm{Nu} = \frac{h}{\Delta T} \left. \dpd{T}{z} \right |_\text{wall} = \frac{1}{\gamma} \left. \dpd{\theta}{z}\right|_\text{wall} - 1.
    \label{rayben:eq:nusselt}
\end{equation}

Rayleigh-Bénard convection has been studied profusely in the literature, in different configurations and regimes \cite{Biringen1990, Bodenschatz_2000, Amati2005, Lohse_2010,Belmonte1994}. Here we just consider a few simulations and typical quantifications of the flow, to illustrate how our numerical method can be used for a set of PDEs different from the simplest case of the incompressible Navier-Stokes equation. Using the \textsc{Specter} code we solve \cref{rayben:eq:velocity,rayben:eq:theta,rayben:eq:incomp} inside a box of size ${L_x \times L_y \times L_z = 2\pi \times 2\pi \times 1}$. Values of the spatial resolution, of the relevant parameters, and of the characteristic dimensionless numbers are given in \cref{rayben:tbl:convection}. It is worth noting that a turbulent convection regime is attained in all the cases.

\begin{figure}
	\centering
	\includegraphics[width=\textwidth,keepaspectratio=true]{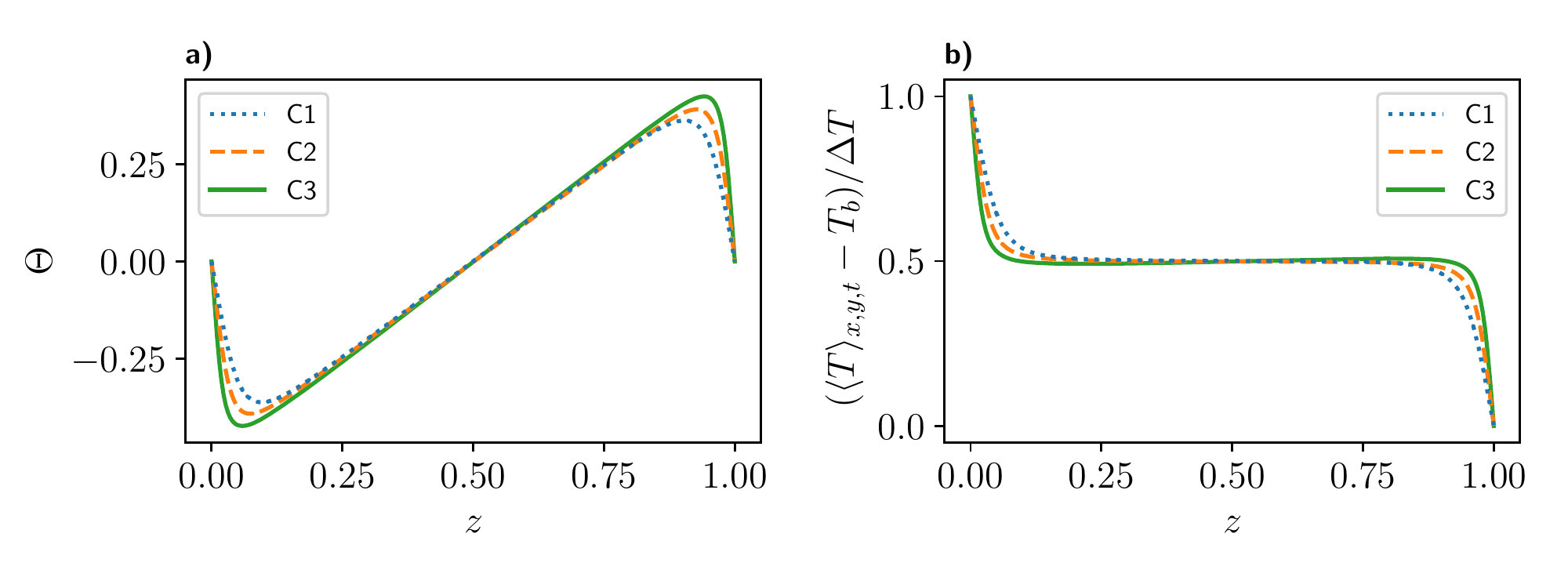}
	\caption{\textbf{a)}: Mean temperature correction (in velocity units) $\Theta$ as a function of the height $z$. The three simulations at different resolutions are shown (see inset for references). \textbf{b)}: Normalized mean total temperature profile $(\expval{T}_{x,y,t}-T_b)/\Delta T$ as a function of the height $z$. Note how temperature becomes uniform in the center of the domain.}
	\label{rayben:fig:rb_profile}
\end{figure}

As for the channel flow, we start by computing the diagonal elements of the two-point correlation tensor. The entries associated with the velocity components are the same as those given in \cref{chan:eq:two_point}, while the temperature dependent elements are defined as
\begin{align}
Q_{\theta j} (z,r_l) = Q_{j\theta} (z,r_l) &= \frac{\expval{\theta'(\bm x) v_j'(\bm x + r_l \bm{\hat r}_{l})}_{x,y,t}}{\expval{\theta'(\bm x) v_j'(\bm x)}_{x,y,t}}, \qquad  \qquad \text{for } j \in \{x,y,z\}, \\
Q_{\theta \theta} (z,r_l) &= \frac{\expval{\theta'(\bm x) \theta'(\bm x + r_l \bm{\hat r}_{l})}_{x,y,t}}{\expval{\theta'(\bm x) \theta(\bm x)}_{x,y,t}}.
\end{align}
As before, primed variables represent fluctuations, i.e., we decompose ${\theta = \Theta + \theta'}$, with capitalized variables denoting mean vertical profiles, i.e., $\Theta = \expval{\theta}_{x,y,t}$. The result of the two-point correlations for simulation C3 is shown in \cref{rayben:fig:rb_two_point} at two different heights, one near the height of maximum turbulent r.m.s.~thermal fluctuations ($z = 0.02$), and the other in the center of the domain. As for the channel flow, horizontal correlations decay to zero before the half box length, and then fluctuate around zero. However, for convection the two-point correlations decay slower than in the channel flow, in good agreement with the development of large-scale convective cells in the flow as will be shown in visualizations later.

Statistically, the convection process should remove heat from the hot plate and transport it to the cold plate. This can be verified by estimating the mean profile for the temperature correction $\Theta$, which should be negative near the bottom plate (i.e., colder than the non-convective solution), and positive near the the top plate (i.e., hotter). In \cref{rayben:fig:rb_profile} we show $\Theta$ for all our simulations. The results are compatible with this observation. Even more, as expected, increasing $\Ra$ (i.e., increasing convective action) results in larger temperature corrections which tend to concentrate closer to the plates. In terms of the temperature correction in velocity units $\theta$, it is straightforward to recover the actual pointwise temperature (normalized by the temperature difference at the boundaries) as
\begin{equation}
\frac{T(x,y,z) - T_b}{\Delta T} = \frac{\theta(x,y,z) - \gamma z}{\gamma h}.
\end{equation}
Using this expression we can calculate the mean temperature profile $(\expval{T}_{x,y,t} - T_b)/\Delta T$, which is also shown in \cref{rayben:fig:rb_profile}. The result is compatible with the profile reported in \cite{Sakakibara2004}, in which temperature is approximately constant in most of the domain  (i.e., convection mixes fluid elements resulting in homogeneous temperature) except in a small region near each plate associated to the thermal boundary layer. As expected, the latter region becomes narrower with increasing $\Ra$.

\begin{figure}
	\centering
	\includegraphics[width=\textwidth,keepaspectratio=true]{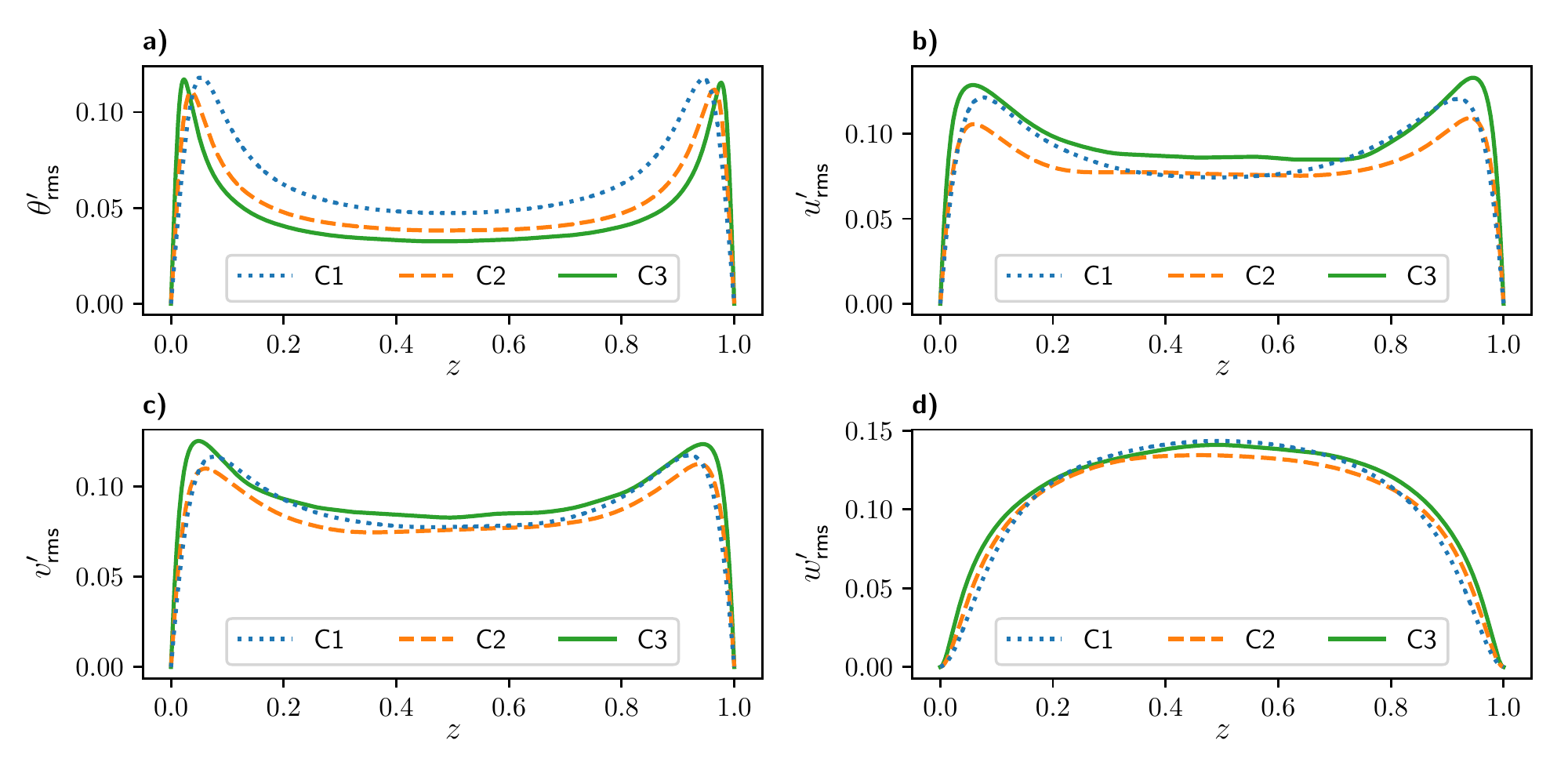}
	\caption{Root mean square (r.m.s.) fluctuations of \textbf{a)}: the turbulent temperature fluctuation in velocity units $\theta'_\text{rms}$, \textbf{b)}: the turbulent velocity component in the $x$ direction $u'_\text{rms}$, \textbf{c)}: the turbulent velocity component in the $y$ direction $v'_\text{rms}$, and \textbf{d)}: the turbulent velocity component in the wall normal direction $w'_\text{rms}$. References for the three simulations are indicated in the insets.}
	\label{rayben:fig:rb_turbulence}
\end{figure}

We now study the variation of r.m.s.~fluctuations as a function of the height, both for the temperature correction and for the velocity. The results are shown in \cref{rayben:fig:rb_turbulence}. The first thing to notice is the symmetry of the profiles, indicating an appropriate timespan for the computed averages, as well as the resemblance between the profiles for $u'_\text{rms}$ and $v'_\text{rms}$, which is expected from the symmetry between $x$ and $y$ directions in the configuration. Additionally, for increasing values of $\Ra$ it is clearly seen that the maximum r.m.s.~fluctuations of $\theta'_\text{rms}$, $u'_\text{rms}$ and $v'_\text{rms}$ tend to concentrate closer to the plates. The increase in the amplitude of their maxima as a function of $\Ra$ is, however, mild, in agreement with results in \cite{Amati2005}. Another important thing to notice is that all $\theta'$, $u'$, and $v'$ attain their maximum r.m.s.~values near the wall, whereas $w'$ is considerably smaller in the same region, which is consistent with results in \cite{Belmonte1994}. On the contrary, it is in the center of the box where $w'$ attains its maximum r.m.s.~amplitude, where it becomes the most relevant quantity. The opposite behavior is found when analyzing $\theta'$, as its r.m.s.~value considerably drops in the center of the box and its centerline value decreases monotonously with $\Ra$, as reported in \cite{Amati2005}.

\begin{figure}
	\centering
	\includegraphics[width=.9\textwidth,keepaspectratio=true]{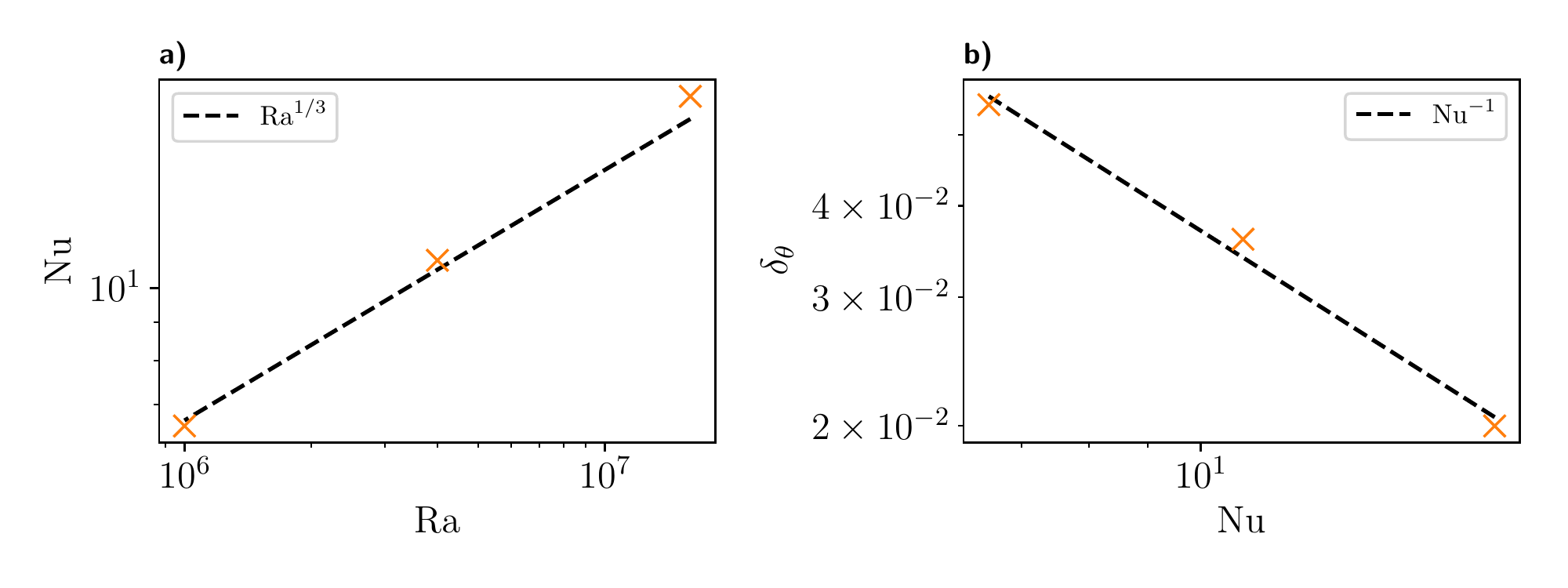}
	\caption{\textbf{a)} Average Nusselt number $\Nu$ as a function of the Rayleigh number $\Ra$ for simulations C1, C2, and C3. A $\Ra^{1/3}$ scaling is shown by the dashed line as a reference. \textbf{b)}: Thermal boundary layer thickness $\delta_\theta$ as a function of the Nusselt number $\Nu$ for simulations C1, C2, and C3, along with a $\Nu^{-1}$ dashed line for reference.}
	\label{rayben:fig:rb_scaling}
\end{figure}

As a way to compare our results with previous simulations of convection, we look at the scaling of the Nusselt number with the Rayleigh number, and at the scaling of the thickness of the thermal boundary layer. Many studies of Rayleigh-Bénard turbulent convection reported a dependence of $\Nu$ (which usually cannot be directly controlled and is estimated \textit {a posteriori}) with $\Ra$ (the control parameter). In simulations in \cite{Amati2005} it was found a scaling $\Nu \sim \Ra^{1/3}$, with the product $\Nu \, \Ra^{1/3}$ being approximately constant for $\Ra$ ranging from $10^6$ to $10^{14}$. To test these results in our simulations, we estimate the mean Nusselt number using \cref{rayben:eq:nusselt}. The results are shown in \cref{rayben:tbl:convection} and in \cref{rayben:fig:rb_scaling}, where $\Nu$ is shown as a function of $\Ra$, with a $\Ra^{1/3}$ scaling law indicated for comparison. The results are compatible with those in \cite{Amati2005}. Even more, the product $\Nu \, \Ra^{1/3}$ is in the range $0.066$--$0.071$ for all our simulations (see \cref{rayben:tbl:convection}), in agreement with  experimental results in \cite{Niemela2000} for convection at $\Pr\approx1$. The scaling of the thermal boundary layer $\delta_\theta$ as a function of $\Nu$ is predicted in \cite{Grotzbach1983,Belmonte1994} to be $\delta_\theta \sim \Nu^{-1}$. We tested this scaling in our simulations by estimating $\delta_\theta$ as the height for which $\theta_\text{rms}$ attained its maximum (we verified that other possible definitions for the width of the thermal boundary layer give similar results). In \cref{rayben:fig:rb_scaling} we show $\delta_\theta$ as a function of Nu, with a $\Nu^{-1}$ scaling law for reference. Our results are in agreement with the aforementioned scaling.

\begin{figure}
	\centering
	\includegraphics[width=.9\textwidth,keepaspectratio=true]{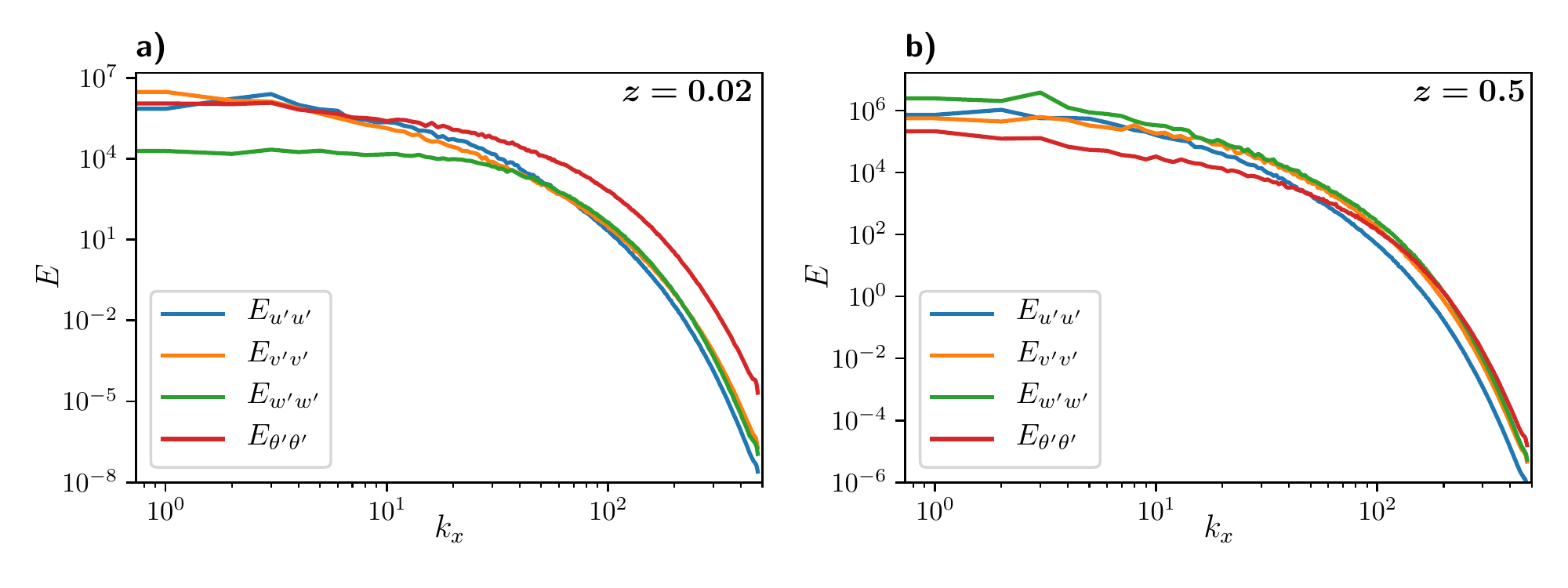}
	\caption{Turbulent energy spectra as a function of the $k_x$ wavenumber for $u'_\text{rms}$, $v'_\text{rms}$, $w'_\text{rms}$, and $\theta'_\text{rms}$ at \textbf{a)}: the thermal boundary layer thickness $z=0.02$, and \textbf{b)}: the center of the domain $z=0.5$ (see insets for references).}
	\label{rayben:fig:rb_spectra}
\end{figure}

As before, we also study the scale dependence of the turbulent fluctuations utilizing the 1D energy spectra, defined for the velocity field in the same way as \cref{chan:eq:spectrax,chan:eq:spectray} and for $\theta$, similarly as
\begin{align}
	E_{\theta'\theta'} (k_x,z) &= \sum_{k_y} \abs{\hat{\theta}'(k_x,k_y,z)}^2,\\
	E_{\theta'\theta'} (k_y,z) &= \sum_{k_x} \abs{\hat{\theta}'(k_x,k_y,z)}^2.
\end{align}
The resulting spectra are shown in \cref{rayben:fig:rb_spectra} as a function of $k_x$ (a similar result is obtained for the spectra as a function of $k_y$) for simulation C3 at two different heights: one at $z=0.02\approx \delta_\theta$, and the other at the center of the box. The first thing to notice is that smooth spectra without notable aliasing are obtained in all cases. Another interesting feature is the absence of a sharp peak for low wavenumbers (i.e., no characteristic roll lengthscale), indicating that the attained regime is indeed of turbulent Rayleigh-Bénard convection. As noted before in \cref{rayben:fig:rb_turbulence}, near the wall and for low wavenumbers, $\theta'$, $u'$, and $v'$ are approximately equal in amplitude, whereas the vertical velocity $w'$ is considerably smaller. This feature is not present for larger wavenumbers, where the three velocity components have approximately the same power and $\theta'$ has significant more power, indicating the presence of very small-scale structures in latter field. The behavior at the center of the box is considerably different. For low wavenumbers it is the vertical velocity $w'$ that contains most of the energy, while $\theta'$ has a smaller power than any velocity component, both facts also in agreement with the features found in \cref{rayben:fig:rb_turbulence}. As the wavenumber increases, nonetheless, the energy becomes more equally distributed between $\theta'$ and the velocity components.

\begin{figure}
	\centering
	\includegraphics[width=\textwidth,keepaspectratio=true]{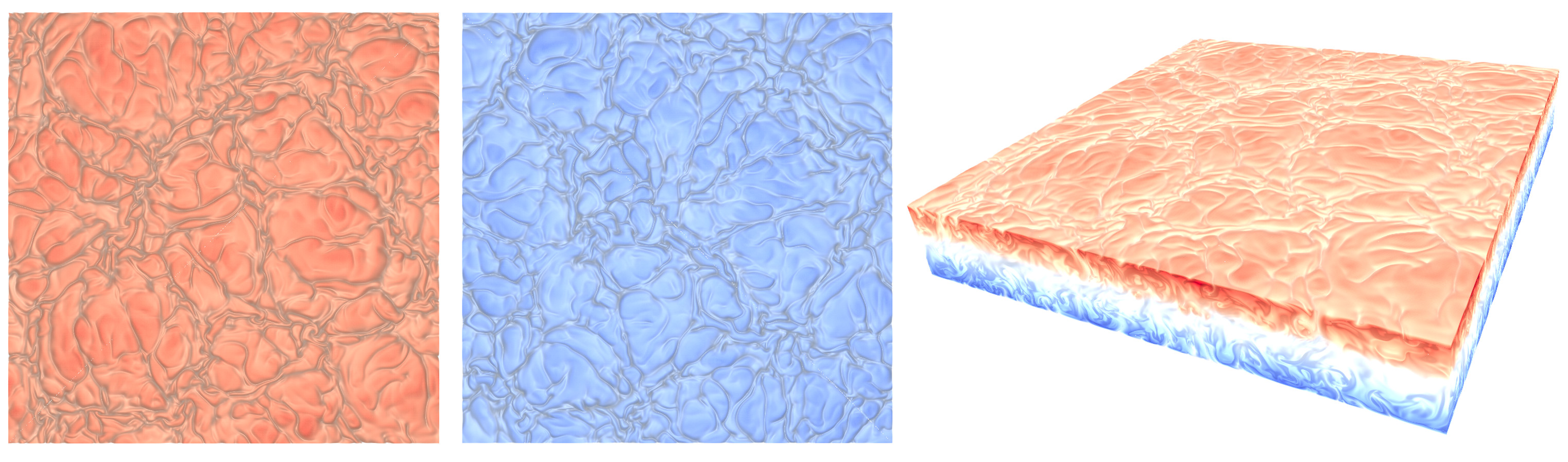}
	\caption{3D rendering of the temperature correction $\theta$ in run C3. Note the total temperature includes the background temperature profile, which is not shown here. From left to right: view of the top of the domain, view from the bottom, and side view. Note hot fluid (in red) is transported to the top while cold fluid (in blue) is transported to the bottom.}
	\label{fig:conv:rendering}
\end{figure}

Finally, \cref{fig:conv:rendering} shows a 3D rendering of the temperature correction $\theta$ using the software \textsc{Vapor}. Note the total temperature is given by $\theta \sqrt{\Delta T/(\alpha g h)}$ plus the background temperature profile, and as a result the renderings only show how convection corrects the background temperature. As expected, hot fluid is transported to the top while cold fluid is transported to the bottom. The multi-scale formation of convective cells, and of turbulent plumes near the walls \cite{Clyne_2007}, can be also clearly seen.

\section{Conclusions}
\label{sec:conclusion}

We presented a Fourier Continuation-based parallel pseudospectral method for incompressible fluids in cuboid non-periodic domains. The method produces dispersionless and dissipationless spatial derivatives with fast spectral convergence inside the domain, and with high order convergence at the boundaries. Thus, the method has no spurious dispersion, or ``pollution," errors that commonly arise in finite differences of finite elements methods. Incompressibility is imposed by solving a Poisson equation for the pressure. As the method is Fourier-based, the Laplace operator for this problem has a diagonal representation in spectral space, and is well-behaved and easy to invert. As a result, solutions of the Poisson equation for the pressure are fast and computationally inexpensive, with the only overhead (compared with pseudospectral methods in periodic domains) of requiring computation of a homogeneous solution to satisfy the required boundary conditions. However, this homogeneous solution can be found analytically and thus generates a minimal overhead.

Being Fourier-based, the method also allows for fast estimation of spectral transforms in $\mathcal{O}(N_i  \log N_i)$ operations in each spatial direction $i$, even when boundary conditions are not periodic. It is compatible with uniform grids (although refined or nested meshes can also be implemented by splitting the domain in multiple subdomains, and matching boundary conditions between these subdomains). But in the case of uniform grids presented here, it allows for explicit time integration with a mild Courant–Friedrichs–Lewy (CFL) condition dominated by the advection term for sufficiently high Reynolds numbers, and thus with a time step for stability that scales linearly with the spatial resolution.

We also presented two time stepping methods, using a time-splitting technique to allow for independent imposition of the boundary conditions for the velocity field (or other fields in the PDEs considered), and for the pressure. A time-splitting forward Euler method was presented that has global error of order $\mathcal{O}(\Delta t^2)$ and that satisfies the boundary conditions with error $\mathcal{O}(\Delta t)$, and we also presented a time-splitting low-storage $o$-th order Runge-Kutta method that has both global and boundary condition errors of $\mathcal{O}(\Delta t^o)$.

The method with the time-splitting Runge-Kutta time evolving scheme was implemented in a publicly available code (\textsc{Specter}), and we briefly described efficient methods for its parallelization. This implementation of the method was validated against two problems with non-periodic boundary conditions: channel flow, and plane Rayleigh-B\'enard convection under  the  Boussinesq  approximation. For channel flows, we also compared our results with previous simulations using other high-order numerical methods. In both cases the  method  yields  results  compatible  with  previous studies.

\section*{Acknowledgements}

The authors acknowledge support from CONICET and ANPCyT through PIP Grant No.~11220150100324CO and PICT Grant No.~2015-3530. This work was also supported by NSF and AFOSR through contracts DMS-1714169 and FA9550-15-1-0043, and by the NSSEFF Vannevar Bush Fellowship under contract number N00014-16-1-2808. We also thank the Physics Department at the University of Buenos Aires for providing computing time on its \textsc{Dirac} cluster.
\appendix
\section{Generation of random solenoidal 3D vector fields}
\label{app:noise}

In order to generate 3D incompressible noise that satisfies the no-slip condition to perturb the velocity field, we use a basis of eigenfunctions of the curl operator introduced in \cite{Clever1997}. These eigenfunctions are obtained from the scalar potentials
\begin{align}
\label{app:eq:phi}
\phi(x,y,z) &= \sum_{nlm} a_{nlm}\exp[i(k_{x,n}x + k_{y,l}y)] g_m(z'), \\
\label{app:eq:psi}
\psi(x,y,z) &= \sum_{nlm} b_{nlm}\exp[i(k_{x,n}x + k_{y,l}y)] \sin \left[ m \pi \left( z'  +\tfrac{1}{2} \right) \right].
\end{align}
by means of the toroidal-poloidal decomposition $\bm q = \bm \nabla \times (\bm \nabla \times \phi \bm {\hat z}) + \bm \nabla \times \psi \bm{\hat z}$. The functions $\bm q$ are eigenfunctions of the curl, and thus generate incompressible flows. Here, $z'$ is the result of mapping $z$ into the $[-1/2,1/2]$ domain, i.e., $z' = z/L_z - 1/2$, and $g_m(z')$ are the Chandrasekhar-Reid eigenfunctions (see pp.~634-637 of Ref.~\cite{Chandrasekhar1981}),
\begin{equation}
g_m(z') = \begin{dcases}
\dfrac{\cosh(\lambda_m z')}{\cosh(\lambda_m /2)} - \dfrac{\cos(\lambda_m z')}{\cos(\lambda_m /2)} &\text{\qquad for $m$ odd,}\\
\dfrac{\sinh(\lambda_m z')}{\sinh(\lambda_m /2)} - \dfrac{\sin(\lambda_m z')}{\sin(\lambda_m /2)} &\text{\qquad for $m$ even,}
\end{dcases}
\end{equation}
where the sequence $\lambda_m$ is constructed from the condition $g_m'(-1/2) = g_m'(1/2) = 0$.

Using \cref{app:eq:phi,app:eq:psi} we can construct a random and incompressible 3D vector by generating a superposition of the $\bm q$ functions with random phases and amplitudes for both the $a_{nlm}$ and $b_{nlm}$ coefficients in the range $4 \le k_{x,n}^2 + k_{y,m}^2 \le 25$ and $2 \le m \le 5$. The amplitudes decay as $k_{nlm}^{-4}$ for $\phi$ and $k_{nlm}^{-3}$ for $\psi$, with $k_{nlm} = (k_{x,n}^2 + k_{y,m}^2 + \lambda_m^2)^{1/2}$. This generates a random 3D incompressible vector field that decays as $k^{-2}$, and whose poloidal and toroidal components are approximately balanced. Note that other choices for the decay of the spectrum of the perturbation can be easily obtained with other choices for the decay of the $a_{nlm}$ and $b_{nlm}$ coefficients.

\section{Mathematical aspects of the FC-Gram Fourier continuation method}
\label{app:FC-Gram}

As mentioned in \cref{sec:FCGram}, and illustrated in \cref{fc:fig:fc_demo}, the Fourier continuation (FC) method utilizes a certain periodic-extension approach, the FC-Gram method, to produce rapidly-convergent Fourier series representations of non-periodic functions defined on one-dimensional intervals. Thus, for a given function $f$, which, without loss of generality, we assume
is defined in the interval $[0,1]$,
\[
f: [0,1] \to \mathbb{R},
\]
the FC method produces a $b$-periodic function
\[
f^c : [0,b]\to \mathbb{R} \quad (b>1),
\]
defined on the interval $[0,b]\supset [0,1]$, which closely approximates $f(x)$ throughout the original interval $[0,1]$---up to and including the endpoints $0$ and $1$.

Following~\cite[Sec. 3.1]{Amlani2016}, more precisely, given a column vector $\mathbf{f} = (f_0,\dots,f_{N-1})^T$ containing point-values of the function $f$ on the equispaced grid
$0 = x_0<x_1<\dots <x_{N-1} =1$, $f_i =f(x_i)$, the FC-Gram method~\cite{Albin2011,Lyon2010a,Amlani2016} uses a subset of the given function values on small numbers $d_\ell$ and $d_r$ of matching points $\{x_0,..,x_{d_\ell-1}\}$ and $\{x_{N-d_r},...,x_{N-1}\}$
contained in small subintervals on the left and right ends of the interval $[0,1]$ (of lengths $\delta_\ell = (d_\ell-1)\Delta x$ and $\delta_r =(d_r-1)\Delta x$, where $\Delta x$ is the distance between matching points) to produce, at first, a discrete periodic
extension. Use of different numbers of matching points $d_\ell$ and $d_r$, $d_\ell\ne d_r$ is desirable, for example, in cases in which one of the interval endpoints corresponds to a point on the boundary
of a computational domain $\Omega$ used for a PDE solution, while the other corresponds to a point interior to $\Omega$---at which the numerical solution is more accurate and whose error is smoother; see,
e.g.,~\cite{Albin2011}. Throughout this paper we have used the values $d_\ell = d_r= d$ with $d=5$ and $d=7$.

In order to obtain the desired discrete periodic expansion, the FC-Gram algorithm appends a number $C$ of continuation function values in the interval $[1,b]$ to the existing function data, so that
the extension transitions smoothly from $f_{N-1}$ back to $f_0$, as depicted in \cref{fc:fig:fc_demo}. The resulting vector $\mathbf{f}^c$ can be viewed as a discrete set of values of a smooth
and periodic function which is suitable for high-order approximation by means of the FFT algorithm in an interval of length $(N+C)\Delta x$.  The $C$ continuation values are produced on the
basis of the discrete function defined by the vector $\mathbf{f}$ together with a translation of it by a distance $b$. In detail, defining the sets $\mathcal{D}_\ell = \{b+x_0, b+x_1, ...,b+x_{d_{\ell-1}}\}$ and $\mathcal{D}_{r} = \{x_{N-d_r}, x_{N-(d_r-1)},..., x_{N-1}\}$, the additional $C$ needed values in the interval $[1,b]$ are obtained as point values of an auxiliary trigonometric polynomial of periodicity
interval $[1-\delta_r, 2b-(1-\delta_r)]$ (with appropriately selected bandwidth) which closely approximates the function values on $\mathcal{D}_r\cup\mathcal{D}_\ell$. This approximating trigonometric
polynomial is obtained as the result of a two-step process, namely: (1) Projection onto bases of orthogonal polynomials (Gram bases), and (2) continuation through use of a precomputed set of continuations-to-zero of each Gram polynomial, as explained in what follows.

The polynomial projection mentioned in step (1) above for the function values on $\mathcal{D}_r$ and $\mathcal{D}_\ell$ (cf.~\cite{Lyon2010a}) relies on use of a basis $\mathcal{B}_r$ (resp. $\mathcal{B}_\ell$), called the Gram basis, of the space of polynomials of degree $<d_r$ (resp. $d_\ell$) on the interval $[1-\delta_r,1]$ (resp. $[b,b+\delta_\ell]$) which is orthonormal with
respect to the discrete scalar product $(\cdot,\cdot)_{r}$ (resp. $(\cdot,\cdot)_\ell$) defined by the discretization points $\mathcal{D}_r$ (resp. $\mathcal{D}_\ell$):
\begin{equation}\label{eq:innerproduct}
(g,h)_{r} = \sum_{x_i \in \mathcal{D}_r} g(x_i)h(x_i),
\end{equation}
with a similar definition for $(g,h)_\ell$. The values of the resulting orthogonal polynomials at the discretization points in $\mathcal{D}_r$ (resp. $\mathcal{D}_\ell$) can be easily obtained by
evaluating the $QR$ factorization of the corresponding Vandermonde matrix. In view of the orthogonality property of the Gram polynomials, any given function can easily be projected onto the polynomial space
directly via scalar product with each one of the orthogonal polynomials---for which, conveniently, only the function values at the {\em Cartesian} discretization points are required. Thus, the use of the Gram polynomial basis makes it easy to produce highly accurate approximations of various orders $r$ of accuracy, by polynomials $p_r$ whose values can be explicitly computed by means of certain well-conditioned linear algebra operations.

The algorithm also utilizes precomputed extensions, one for each polynomial in the Gram basis, into a smooth function defined for $x\geq 1-\delta_r$ which approximates $p_r$ closely in the matching interval $[1-\delta_r,1]$, and which blends smoothly to zero for
$x\geq b.$ The rightward extensions, for example, are constructed as appropriately oversampled least squares approximations by Fourier series of periodicity interval $[1-\delta_r,
2b-(1-\delta_r)]$. Utilizing such smooth blending functions the algorithm proceeds to step (2): The evaluation of an extension from the function values at the set of points
$\mathcal{D}_r\cup\mathcal{D}_\ell$. This is achieved, simply, by projection of the given set of function values onto the polynomial basis, followed by extension via the aforementioned rightward and
leftward extension of Gram polynomials. With the extension in hand, an application of the discrete Fourier transform on the interval $[0,b]$ to the vector of function values $\mathbf{f}$ augmented by the $C$ ``continuation'' values yields the desired trigonometric polynomial
\begin{equation}
\label{eq:fcseries}
f^c(x) = \displaystyle\sum_{k=-M}^{M} a_k e^{(2\pi i k x)/b} \quad \text{s.t.}~f^c(x_i) = f(x_i),~i=0,...,N-1.
\end{equation}
For efficiency, the discrete Fourier transform is implemented by means
of the Fast Fourier Transform (FFT).

The resulting continuation operation can be expressed in a block matrix form as
\begin{equation}\label{eq:FCoperator}
\mathbf{f}^c = \begin{bmatrix} I\\A\end{bmatrix} \mathbf{f} = \begin{bmatrix} \mathbf{f} \\ A\mathbf{f}\end{bmatrix},
\end{equation}
where $\mathbf{f}^c$ is a vector of the $N+C$ continued function values, $I$ is the $N\times N$ identity matrix and $A$ is the matrix containing the blend-to-zero continuation information. Defining the vector of matching points for the left and right as 
\begin{equation}\label{eq:flfr}
\mathbf{f}_\ell = \left(f_0, f_1, ..., f_{d_\ell-1}\right)^T, \quad  \mathbf{f}_r = \left( f_{N-d_r}, f_{N-d_r+1},..., f_{N-1}\right)^T,
\end{equation}
%\begin{equation}\label{eq:flfr}
%\mathbf{f}_\ell = \begin{pmatrix} f_0 \\ f_1 \\ \vdots \\ f_{d_\ell-1}\end{pmatrix}, \quad  \mathbf{f}_r = \begin{pmatrix} f_{N-d_r} \\ f_{N-d_r+1} \\ \vdots \\ f_{N-1}\end{pmatrix},
%\end{equation}
the matrix $A$ can be expressed in the form \begin{equation}\label{eq:Af}
A \mathbf{f} = A_\ell Q_\ell^T \mathbf{f}_\ell + A_r Q_r^T \mathbf{f}_r,
\end{equation}
where the columns of $Q_\ell$ and $Q_r$ contain the $d_\ell, d_r$ point values of each element of the corresponding Gram polynomial basis, and where the columns of $A_\ell$ and $A_r$ contain the
corresponding $C$ values that blend the polynomials in the left and the right Gram bases to zero. This step, which is responsible for all the ill conditioning in the continuation problem, can advantageously
be performed as a precomputed operation, in high-precision accuracy, to produce a small re-usable set of parameters (a set of numbers proportional to both the number of Gram polynomials and
extension points used---e.g., $231$ numbers in our case), which can be utilized for
continuation of arbitrary functions with negligible ill conditioning; full details in these regards may be found, e.g., in~\cite[Sec. 3.1]{Amlani2016}.
\vskip 1cm

\bibliographystyle{elsarticle-num-names}
\bibliography{spectre}

\end{document}